\documentclass[superscriptaddress, preprint, amsmath, showkeys, a4paper]{revtex4-1} 
\usepackage{subfig, multirow, rotating,xcolor}
\usepackage[]{graphicx}
\usepackage[section]{placeins}
\usepackage[font=small,justification=justified,singlelinecheck=false]{caption}
\usepackage{xr-hyper}
\usepackage[driverfallback=dvipdfm]{hyperref}
\usepackage{cleveref}
\usepackage{accents}
\usepackage[normalem]{ulem}
\renewcommand\eqref[1]{(\ref{#1})}
\newcommand\figref[1]{Figure~\ref{#1}}

\newcommand{\ac}{a^{\dagger}}
\newcommand{\cc}{c_{k}^{\dagger}}
\newcommand{\aan}{a}
\newcommand{\ca}{c_{k}}
\newcommand{\bc}{b_{q}^{\dagger}}
\newcommand{\ba}{b_{q}}
\newcommand{\omegq}{\omega_{q}}
\newcommand{\ek}{\varepsilon_{k}}
\newcommand{\ea}{\varepsilon_{a}}
\newcommand{\ene}{\varepsilon}
\newcommand{\eatil}{\tilde{\varepsilon}_{a}}

\newcommand{\lsb}{\left[}
\newcommand{\rsb}{\right]}
\newcommand{\lcb}{\left\{}
\newcommand{\rcb}{\right\}}

\begin{document}
\title[]{Time-dependent electron transfer and energy dissipation in condensed media}
\author{Elvis F. Arguelles}
\email[]{arguelles@issp.u-tokyo.ac.jp}
\author{Osamu Sugino}
\email[]{sugino@issp.u-tokyo.ac.jp}
\affiliation{Institute for Solid State Physics, The University of Tokyo, 5-1-5, Kashiwanoha, Kashiwa, Chiba 277-8581, Japan}

\begin{abstract}
	We study a moving adsorbate interacting with a metal electrode immersed in a solvent using
	the time-dependent Newns-Anderson-Schmickler model Hamiltonian. 
	We have adopted a semiclassical trajectory treatment of the adsorbate 
	to discuss the electron and energy transfers that occur between the adsorbate and the electrode.
	Using the Keldysh Green's function scheme, we found a non-adiabatically suppressed electron transfer 
	caused by the motion of the adsorbate and coupling with bath phonons that model the solvent. 
	The energy is thus dissipated into electron-hole pair excitations, 
	which are hindered by interacting with the solvent modes and facilitated by the applied electrode potential. 
	The average energy transfer rate is discussed in terms of the electron friction coefficient 
	and given an analytical expression in the slow-motion limit.
\end{abstract}

\date{\today}

\maketitle
\section{Introduction}
Understanding the fundamental concept of electron transfer has been the subject 
of intense theoretical research over the past decades due to its importance in a wide-range of fields
including  physics, chemistry and biology.
One of the most widely studied systems is that of atoms or molecules (adsorbates) adsorbed 
on a metal electrode
immersed in a solution.
Electron transfer between the adsorbate and the electrode forms the basis 
of many technologically important electrochemical reactions.
It is intrinsically coupled to the vibrational 
degrees of freedom of the solvent environment, often represented by a phonon bath. 

Since the pioneering semiclassical\cite{Marcus1965,Hush1958}
and quantum mechanical\cite{Levich1970,Schmickler1986} studies, the theory of electrochemical electron transfer has
progressed over the years.
Research on this topic can generally be divided into two categories, 
depending on the strength of the electronic coupling or the size of the electronic transfer integral. 
In the strong limit, the characteristic time scales of the adsorbate 
and the electrode are shorter than those of the solvent, allowing electron transfer 
to proceed adiabatically\cite{Schmickler1986,Marcus1965,Hush1958}.
In the weak limit, on the other hand, electron transfer occurs non-adiabatically 
\cite{Levich1970}. Here, 
"strong" means that the interactions are large enough to establish equilibrium between
the adsorbate and the electrode, but are generally weaker relative 
to the electron-phonon (e-ph) interaction. 
This is reminiscent of the polaronic system\cite{Hewson1974,Citrin1977} wherein it was
argued that the energy shift in proportion to 
the e-ph couplings must be greater than the electronic couplings
to be detected by photoemission experiments.

In recent non-adiabatic formulations\cite{Song1993,Sebastian1989,Smith1993,Mohr2000,Tanaka1999},
the restrictions on the strength electronic interactions have been substantially relaxed,
allowing for all ranges to be considered.
Most theories using time-independent
Hamiltonians instead ignore non-adiabatic effects induced by moving adsorbates. 
In this context, electron transfer rates are calculated at a certain fixed
nuclear position of the adsorbate, where electrons have relaxed to
their ground state, or in short, within the Born-Oppenheimmer picture.
This approximation is known to break down even for slow reactant velocities corresponding to the thermal energy,
and more so when dealing with metal electrodes where any dynamical processes would 
lead to electron-hole (e-h) pairs excitations\cite{Wodtke2004}. 

It is crucial at this stage to distinguish between the concepts 
of non-adiabaticity in the context of both time-independent and moving adsorbate approaches.
In the former scenario, non-adiabaticity arises from weak electronic interactions 
in chemical reactions. In models utilizing potential energy surfaces (PES),
the electronic interactions are commonly described by the coupling between two diabatic 
PES\cite{Piechota2019}. In the latter case, the motion of the 
adsorbate impedes the relaxation of electrons to the adiabatic ground state,
meaning that the electron states become time-dependent and non-adiabatic.
Henceforth, we will focus on non-adiabaticity in the context of adsorbate motion.

The charge transfer dynamics on metal substrates is a well studied topic in surface physics mostly on the basis of
the time-dependent Newns-Anderson model Hamiltonian\cite{Brako1981,Blandin1976,Yoshimori1984,Kasai1987,Yoshimori1986,Mizielinski2005}.
Within this model, the time-dependence of adsorbate energy level and the electronic coupling due to the
motion of the adsorbate are explicitly considered, yielding many rich and interesting results. 
For one, the non-adiabatic adsorbate orbital occupancy deviates significantly from the adiabatic value 
during its encounter with the metal substrate. Further, the energy dissipation through e-h
pairs\cite{Brako1980,Brako1981,Plihal1998,Mizielinski2005} and vibrational\cite{Newns1986,Kasai1991,Gross1993} excitations
can naturally be included in the model. 
While the vibrational non-adiabaticity have been previously considered in electrochemical systems
\cite{Lam2019}, the time-dependent electron transfer as well as the effects of continuum of electronic
states of the metal electrodes have so far received very little attention. 

In the present work, we explore these
non-adiabatic effects by employing a time-dependent version of the model Hamiltonian 
originally introduced by Schmickler\cite{Schmickler1986}.
Using the Keldysh formalism, we derive the time-dependent adsorbate orbital occupancy from which
the electron and  non-adiabatic energy transfer rates can be obtained. The effects of
adsorbate velocity and the strength of electron-bath phonon coupling on the electron and energy
transfer rates are investigated by numerical calculations. 
By considering the limit of static adsorbate and high temperature, our formulations are reduced
to the Marcus theory (Appendix \ref{appendix:marcus}). Furthermore, in the limit of slow 
adsorbate motion, we derive the analytical expressions of the electronic friction coefficient and
average energy transfer rate. We discuss the effects
of adsorbate electron-solvent modes coupling and electrode potential to the energy exchange and electronic
friction. These expressions are then rederived from 
the e-h excitation probability of the metal electrons, from which, we discuss the adsorbate
sticking probability. This work is organized as follows. In \ref{section:Model}, we present the
theoretical model that describes the essential physics of electron transfer in electrochemical
systems. This is followed by the derivations of the non-adiabatic orbital occupancy and 
and energy transfer rate in \ref{section:Theory}. We specialize on the slow motion limit
and obtain an analytical expression of the electronic friction coefficient both from 
the nearly adiabatic expansions of the time-dependent parameters and e-h excitation
probability. In \ref{section:Application}, we apply the formalism to the electrochemical proton coupled electron
transfer reaction where we present some illustrative numerical calculations of the orbital occupancy,
and electron and energy transfer rates. Thereupon, we discuss
the e-h excitation probability and estimate the sticking probability in the high
temperature limit. Finally, we summarize our results and present our conclusions in \ref{section:summary}.

\section{Model}
\label{section:Model}
\begin{figure}
\centering
\captionsetup{justification=raggedright,singlelinecheck=false}
\includegraphics[scale=0.5]{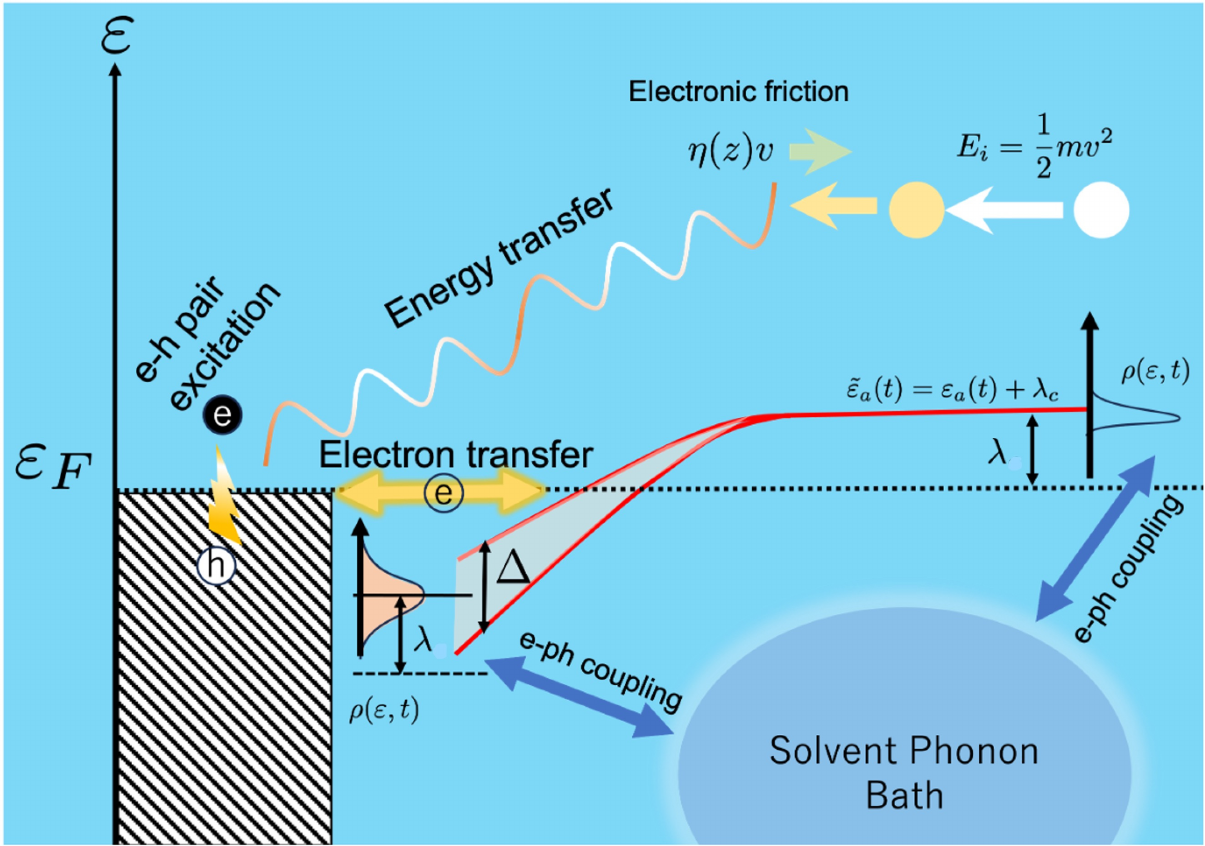}
\caption{\label{fig:system} Schematic diagram of the time-dependent electron transfer
and energy exchange in an electrochemical system. An adsorbate with velocity $v$ and initial kinetic
energy $E_{i}$ located far from the metal electrode is characterized by
a sharp resonance $\rho(\ene,t)$ and an orbital energy level $\tilde{\ene}_{a}(t)$
renormalized by the interaction with the solvent modes. The renormalization shifts the orbital energy
level above the Fermi level ($\ene_{F}$) by the reorganization energy $\lambda$ which quantifies the electron-solvent
mode coupling.
As the adsorbate moves closer to the surface, $\rho(\ene,t)$ is broadened with a width
$\Delta(t)$ due to the electron transfer from the surface. This broadening is accompanied
by the occupation of the adsorbate orbital depicted by $\tilde{\ene}_{a}(t)$ crossing
$\ene_{F}$. In addition, the moving adsorbate induce electron-hole excitations in the metal
surface which in turn induce a frictional force $\eta(z)v$ that slows it down.}
\end{figure}

The electrochemical system we wish to investigate consists of a moving adsorbate
and a metal electrode solvated in a solution. The electron transfer and energy exchange
mechanisms of this system are schematically represented in \figref{fig:system}. 
We assume that the system is in equilibrium initially 
and the adsorbate is located sufficiently far away from the metal electrode 
so that any electron tunneling is impossible. The ions and atoms comprising the solvent 
are assumed to move infinitesimally in such a way that they can be represented by 
harmonic vibrations with infinite modes that form a phonon bath. 
At distances far from the surface of the electrode, 
the adsorbate orbital is characterized by a sharp resonance near the Fermi level ($\ene_{F}$).
As the adsorbate approaches the 
electrode, this state broadens as the energy level crosses $\ene_{F}$ and becomes filled. 
The broadening is proportional to the square of the electronic overlap integrals between the metal and the 
adsorbate, represented by the resonance width $\Delta$. In addition, we assume that the adsorbate electron
couples with the harmonic solvent modes linearly and shifts the energy level by an amount 
equivalent to the reorganization energy $\lambda$, a quantity proportional
to the e-ph interaction. Furthermore, we assume that the adsorbate possesses a kinetic energy initially
which is dissipated towards the excitation of e-h pairs in the metal
electrode. These excitations consequently induce an electronic frictional force that slows down the
adsorbate. There are of course other possible energy dissipation channels. However, in the present work,
we are interested in non-adiabatic effects that are unique to a system involving a metal electrode.

In electrochemical systems, the electronic interaction between the
metal and adsorbate is generally weak and only one spin state is typically occupied.
Therefore, the electron repulsion interaction is neglected in the following. Further, to simplify our calculations,
we invoke the so-called $trajectory~approximation$ which assumes that
the motion of the adsorbate and the dynamics of the system are separated out.
Simply put, the classical trajectory $R(t)$ of the adsorbate is known from the beginning.
In most cases, this approximation works well for adsorbates with large masses or moving slowly.
Therefore, our work treats the electrons and phonons quantum mechanically, while the adsorbate nuclear
dynamics classically. This is justified when the thermal energy is larger than the frequency corresponding to the
adsorbate nuclear motion (i.e., $k_{B}T > \hbar\omega_{a}$)\cite{Dou2020}, which
we assume to be the case here.

The total Hamiltonian of the system described above 
may be represented by a time-dependent version of the Newns-Anderson-Schmickler 
model\cite{Schmickler1986} ($\hbar=1$),
\begin{equation}
	\begin{aligned}
		H(t) = & \ea(t)n_{a} + \sum_{k}\ek\cc\ca + \sum_{k}\lsb V_{ak}(t)\ac\ca + H.c. \rsb \\
	+ & \sum_{q}\omegq \bc\ba + (Z-n_{a})\sum_{q}\lambda_{q}\omegq(\bc + \ba),
\end{aligned}
	\label{eq:totham}
\end{equation}
where $n_{a}=\ac a$. Here, $\ea(t)=\ea\lsb R(t) \rsb$ is the time-dependent energy level of the adsorbate
whose Fock space is spanned by its electron annihilation and creation operators,
$a$ and $\ac$, respectively. $\ek$ is the band energy of the non-interacting electrons with wavevector $k$ 
of the metal electrode filled up to the Fermi level, and 
$\ca(\cc)$ are their corresponding annihilation(creation) operators.
$V_{ak}(t)=V_{ak}\lsb R(t) \rsb$ is the overlap integral between
the metal $|k\rangle$ and moving adsorbate $|a\rangle$ electron states, whose time-dependence is assumed to
be separable through $V_{ak}(t)=V_{ak}u(t)$\cite{Yoshimori1986} where, $u(t)$ 
is some time-dependent function defined in the time domain corresponding to the chosen
trajectory $R(t)$. In surface physics literatures, 
there are several choices for the form of $u(t)$ and here for simplicity we take it 
as $u(t) = e^{-\gamma |t|^{2}}$. For a one-dimensional trajectory, $\gamma=\alpha v^{2}$
where $\alpha$ is the decay constant of the surface wave function and $v$ is the
speed of the adsorbate. This choice reflects the fact that empirically, $V_{ak}(t)\propto S$
where $S$ is the overap between the surface wave function $\psi_{k}(r)$ and the adsorbate orbital
$\phi_{a}(r-R(t))$. $\omegq$ is the frequency 
of the harmonic oscillators describing the solvent and $\ba(\bc)$ is
the annihilation(creation) operator of the phonon of mode $q$. The harmonic displacements
$\bc+\ba$ couple linearly to the adsorbate electrons, $n_{a}$ with coupling strength, $\lambda_{q}$
and $Z$ is the charge of the oxidized adsorbate state.

We eliminate the e-ph
coupling term in \eqref{eq:totham} by performing a nonperturbative canonical (Lang-Firsov) transformation
$\tilde{H}=e^{S}He^{-S}$ where $S=\sum_{q}\lambda_{q}n_{a}(\bc-\ba)$. This results in the 
dressing of the adsorbate electron states as $\tilde{a}=ae^{-\sum_{q}\lambda_{q}(\bc-\ba)}\equiv aX$,
$\tilde{a}^{\dagger}=\ac e^{\sum_{q}\lambda_{q}(\bc-\ba)}\equiv \ac X^{\dagger}$, and the shifting of the phonon operators
$\tilde{b}_{q} = \ba -\lambda_{q}n_{a}$, $\tilde{b}^{\dagger}_{q} = \bc - \lambda_{q}n_{a}$ due
to the charging effect. The transformation leaves the metal electron states unchanged.
The transformed Hamiltonian therefore reads
\begin{equation}
	\begin{aligned}
		\tilde{H}(t) = & \eatil(t)n_{a} + \sum_{k}\ek\cc\ca + \sum_{k}\lsb \tilde{V}_{ak}(t)\ac\ca + H.c. \rsb \\
		+ & \sum_{q}\omegq \lsb \bc\ba+Z\lambda_{q}(\bc+\ba) \rsb,
\end{aligned}
	\label{eq:lf_totham}
\end{equation}
where 

\begin{equation}
		\eatil(t)=\ea(t)+(2Z-1)\lambda
	\label{eq:renorm_elevel}
\end{equation}
is the adsorbate orbital energy level renormalized by e-ph
coupling through the reorganization energy defined as $\lambda\equiv\sum_{q}\lambda_{q}^{2}\omegq$; 
$\tilde{V}_{ak}(t)=V_{ak}(t)X^{\dagger}$ is the dressed electronic overlap integral.
As in the local polaron problem\cite{Hewson1980}, $V_{ak}$ is usually
smaller than $\lambda_{q}$ and as a consequence, $\lambda$.  It is therefore sufficient to replace $X^{\dagger}$ with
its expectation value\cite{Chen2005} so that 
$\tilde{V}_{ak}(t) \approx V_{ak}(t)\langle X^{\dagger} \rangle =V_{ak}(t)\exp\lsb \sum_{q}\lambda_{q}^{2}(N_{q}+1/2) \rsb$,
where $N_{q}=\frac{1}{e^{\beta\omegq}-1}$ corresponds to the phonon population and $\beta=1/k_{B}T$.

\section{Theory}
\label{section:Theory}
\subsection{Time-dependent electron transfer}
\label{subsection:et}
To describe the electron transfer, we require the calculation of the time-dependent adsorbate orbital 
occupancy which can be obtained from
Keldysh Green's functions through $\langle n_{a}(t) \rangle = \mathrm{Im}[G^{<}(t,t)]$.
Our starting point is the nonequilibrium Dyson equation of the Keldysh lesser Green's function 
\begin{equation}
	G^{<}(t,t') = G_{0}^{<}(t,t') + \int_{C} dt_{1} \int_{C} dt_{2} G_{0}^{<}(t,t_{1}) \Sigma^{<}(t_{1},t_{2})G^{<}(t_{2},t')
	\label{eq:dysoncontourGless}
\end{equation}
where $C$ is the time contour, $G^{<}_{0}$ is the unperturbed lesser Green's function, and $\Sigma^{<}$ is the self-energy
functional. 
To this end, the contour ordered Green's function
of the nonequilibrium system
is expressed in a form that allows for a diagrammatic perturbation expansion by virtue
of the Wick's theorem. This is achieved by performing a series of transformations to arrive at
an expression governed by the solvable, quadratic unperturbed Hamiltonian and a corresponding 
equilibrium density matrix. By setting the reference time $t_{0} \rightarrow -\infty$, 
and assuming that the initial interactions can be ignored, the contour $C$ can be identified 
as the Keldysh contour\cite{Jauho1994} that begins and ends at $-\infty$.
The contour integration 
in \eqref{eq:dysoncontourGless} is then converted to the real time integration by performing an 
analytical continuation using 
Langreth rules\cite{Langreth1976} resulting in
\begin{equation}
	G^{<} = (1+G^{r}\Sigma^{r})G_{0}^{<}(1+G^{a}\Sigma^{a}) + G^{r}\Sigma^{<}G^{a},
	\label{eq:lang_glessiter}
\end{equation}
where $G^{r(a)}$ is the retarded(advanced) Green's function, $\Sigma^{r(a)}$ is the retarded(advanced) 
self-energy, $\Sigma^{<}$ is the lesser self-energy and $G_{0}^{<}$ is the unperturbed lesser
Green's function.
The first term is identically zero\cite{Odashima2017} within the wide-band approximation, which
we shall extensively exploit in the following.
In integral representation, $G^{<}$ may be written as\cite{Jauho1994}
\begin{equation}
	G^{<}(t,t') = \int_{t_{0}}^{t} dt_{1} \int_{t_{0}}^{t} dt_{2} G^{r}(t,t_{1})\Sigma^{<}(t_{1},t_{2})G^{a}(t_{2},t'). 
	\label{eq:gless}
\end{equation}
The lesser self-energy appearing in \eqref{eq:gless} can be expressed in terms of the 
the renormalized electronic overlap integrals, $\Sigma^{<}(t_{1},t_{2}) = \sum_{k} \tilde{V}_{ak}(t_{1}) g^{<}_{k}(t_{1},t_{2}) \tilde{V}^{*}_{ak}(t_{2})$
with $g^{<}_{k}(t_{1},t_{2}) =   i f(\ek)\exp\left[-i\ek(t_{1}-t_{2}) \right]$ as the unperturbed Green's function 
of the metal electrode electrons and $f(\ek)$ is their corresponding Fermi-Dirac distribution function.
In the wide-band limit, the time-independent part of the electronic overlap integrals can be expressed in terms of the adsorbate level resonance width
which we assume to be energy-independent
through $\Delta = \pi|V_{ak}|^{2} \rho(\ek)$, with $\rho(\ek)$ is the density of states of the metal electrons.
Changing the summation over $k$ to an integral over $\varepsilon$, the lesser self-energy takes the form
\begin{equation}
		\Sigma^{<}(t_{1},t_{2})  = i\int d\ene f(\ene) \sqrt{\frac{\Delta(t_{1})}{\pi}}\sqrt{\frac{\Delta(t_{2})}{\pi}} \
					\exp\left[-i\ene(t_{1}-t_{2}) \right]
	\label{eq:les_selfen2}
\end{equation}
where, $\Delta(t) =  \Delta u(t)^{2}$.
Substituting \eqref{eq:les_selfen2} into \eqref{eq:gless} yields
\begin{equation}
	G^{<}(t,t') = i\int d\ene f(\ene) \int_{t_{0}}^{t} dt_{1}\sqrt{\frac{\Delta(t_{1})}{\pi}} \
	\int_{t_{0}}^{t} dt_{2}\sqrt{\frac{\Delta(t_{2})}{\pi}} G^{r}(t,t_{1})\exp\left[-i\ene(t_{1}-t_{2}) \right]G^{a}(t_{2},t'). 
	\label{eq:gless1}
\end{equation}
As in the polaron systems, we assume that the retarded Green's function can be decoupled as\cite{Zhu2003}
\begin{equation}
	\begin{aligned}
		G^{r(a)}(t,t') = & \mp i\theta(\pm t \mp t')\langle \left\{ \tilde{a}(t),\tilde{a}^{\dagger}(t') \right\} \rangle_{el} \
			\langle X(t)X^{\dagger}(t')\rangle_{ph} \\
			= & ~\tilde{G}^{r(a)}(t,t')\langle X(t)X^{\dagger}(t')\rangle_{ph}.
	\end{aligned}
	\label{eq:gretad}
\end{equation}
Here, $X(t) = \exp\lsb-\sum_{q}\lambda_{q} \left(\bc e^{i\omega t} -\ba e^{-i\omegq t} \right) \rsb$,
and $\langle \ldots \rangle_{e(ph)}$
is the expectation value with respect to electrons(phonons). 
The solvent correlation function $\langle X(t)X^{\dagger}(t') \rangle_{ph} = \mathrm{Tr}\lsb\rho_{ph} X(t)X^{\dagger}(t')\rsb \equiv B(t-t')$ 
is given by
\begin{equation}
	B(t) = \exp\lcb -\sum_{q}\lambda_{q}^{2} \
	\lsb (N_{q}+1)(1-e^{-i\omega_{q}t})+N_{q}(1-e^{i\omega t}) \rsb \rcb.
	\label{eq:bathcor}
\end{equation}
We assume that the solvent bath phonons are characterized by a spectral density
$J(\omega)=\sum_{q=1}^{N}\lambda_{q}^{2}\omega_{q}^{2}\delta(\omega-\omega_{q})$.
Replacing the $q$-mode summation with an integration over $\omega$ and noting the antisymmetric
property of the spectral density, we rewrite \eqref{eq:bathcor} as
\begin{equation}
	B(t) = \exp\lsb \int_{-\infty}^{\infty}d\omega J(\omega) \
	\frac{e^{i\omega t}}{e^{\beta\omega}-1}  \rsb.
	\label{eq:bathcor1}
\end{equation}
We choose the Drude-Lorentz form of the spectral density 
\begin{equation}
	J(\omega) = \frac{2\lambda}{\omega_{0}^{2}}\frac{\gamma\omega}{\gamma^{2}+\omega^{2}},
	\label{eq:sd}
\end{equation}
where $\omega_{0}$ is characteristic frequency of 
the solvent, and $\gamma$ is the cutoff frequency. $\lambda$ is typically in the range of
$0.5-1.0$ eV\cite{Schmickler1986}.
The $\omega$ integration
can be performed analytically by contour integration giving 
\begin{equation}
	B(t) = \exp\lsb \sum_{k=0}^{\infty}C_{k}e^{-\nu_{k}t} \rsb,
	\label{eq:bathcor2}
\end{equation}
where $C_{k=0}=\frac{\lambda\gamma}{\omega_{0}^{2}}\lsb\cot\left(\frac{\beta\gamma}{2}\right)-i\rsb$,
$C_{k\ge1}=4\frac{\lambda\gamma}{\omega_{0}^{2}}\frac{\nu_{k}}{\beta\left(\nu_{k}^{2}-\gamma^{2}\right)}$,
$\nu_{0}\equiv\gamma$, and $\nu_{k}=2\pi k/\beta$ is the $k$th Matsubara frequency.
In the long time (adiabatic) limit, $B(t)\rightarrow 1$.
The Matsubara-decomposed bath correlation function of the form \eqref{eq:bathcor2}
valid for a wide-range of temperature
is frequently used in 
open quantum systems\cite{Tanimura1989,Tanimura1990,Tanimura2020,Lambert2023}.
We next write the Dyson equation for $\tilde{G}^{r}$ as 
\begin{equation}
	\tilde{G}^{r}(t,t_{1}) = G_{0}^{r}(t,t_{1}) + \int d\tau \int d\tau' G_{0}^{r}(t,\tau)\Sigma^{r}(\tau,\tau')\tilde{G}^{r}(\tau',t_{1}) 
	\label{eq:grdyson}
\end{equation}
where the retarded self-energy is defined as
\begin{equation}
	\begin{aligned}
	\Sigma^{r}(\tau,\tau') = & \sum_{k} \tilde{V}_{ak}(\tau) g_{k}^{r}(\tau,\tau') \tilde{V}^{*}_{ak}(\tau')  \\ 
	= & -i \int \frac{d\ene}{\pi} \Delta u(\tau) u(\tau')\exp\left[-i\ene(\tau-\tau') \right] \theta(\tau-\tau') \\
	= & -i \Delta(\tau)\delta(\tau-\tau').
	\end{aligned}
	\label{eq:retselfe}
\end{equation}
In the above expression, we used the wide band approximation and the definition of the unperturbed retarded Green's function of the metal electrode 
$g_{k}^{r}(\tau,\tau')= -i\theta(\tau-\tau') \
\exp\left[-i\ek(\tau-\tau') \right]$ and $\bar{V}_{ak}$.
With \eqref{eq:retselfe} the integral equation \eqref{eq:grdyson} can now be solved\cite{Langreth1991,Wingreen1989} yielding
\begin{equation}
	\tilde{G}^{r}(t,t_{1}) = -i\theta(t-t_{1}) \exp\left\{-i\int_{t_{1}}^{t} d\tau \left[\tilde{\ene}_{a}(\tau) -i\Delta(\tau) \right] \right\},
	\label{eq:gr}
\end{equation}
where $\tilde{\ene}_{a}(t)$ is given by \eqref{eq:renorm_elevel}.
The advanced Green's function can also be obtained by following a similar procedure.
Inserting \eqref{eq:gr} and \eqref{eq:gretad} into \eqref{eq:gless1}, 
the time-dependent occupation number $\langle n_{a}(t) \rangle = \mathrm{Im}[G^{<}(t,t)]$ may be written as
\begin{equation}
	\langle n_{a}(t) \rangle =  \int d\ene f(\ene) \left|p(\ene,t)\right|^2 ,
	\label{eq:occu}
\end{equation}
where
\begin{equation}
	p(\ene,t) =  \int_{t_{0}}^{t} dt'\sqrt{\frac{\Delta(t')}{\pi}} \
	B(t')\exp\left\{-i\int_{t'}^{t} d\tau \left[\tilde{\ene}_{a}(\tau) -i\Delta(\tau) -\ene \right]\right\}.
	\label{eq:p_et}
\end{equation}
The above expression can also be derived from the Heisenberg's equations of motion approach
as presented in Appendix \ref{appendix:eom}. We notice that $|p(\ene,t)|^{2}$ 
may be interpreted as the time-dependent projected
density of states (PDOS) of the adsorbate in electrochemical systems. 
The adiabatic expression of the occupation
can be derived by ignoring the time dependence of $\Delta$ and $\tilde{\ene}_{a}$.
The exponential function may be replaced by
$\exp\left\{-i\left[\tilde{\ene}_{a} -i\Delta -\ene \right](t-t')\right\}$
and the analytical integration of \eqref{eq:p_et} gives the adiabatic PDOS
\begin{equation}
	|p(\ene,t)|^{2} \rightarrow \rho^{ad}(\ene)\equiv \frac{1}{\pi}\frac{\Delta}{\Delta^{2}+\left[ \tilde{\ene}_{a}-\ene\right]^{2}},
	\label{eq:adpdos}
\end{equation}
which consequently yields the adiabatic orbital occupancy as
\begin{equation}
	\langle n^{ad}_{a} \rangle=\int d\ene f(\ene)\rho^{ad}_{a}(\ene).
	\label{eq:nad}
\end{equation}
It is more physically meaningful to work in
terms of the adsorbate position $R(t)$ rather than time. For simplicity, we neglect any motion
parallel to the metal surface and assume that the 
adsorbate moves at a constant velocity $v$, such that 
the one-dimensional trajectory is given by $R(t)\approx z(t)=v|t|$.  
To describe scattering, one may take $t<0$ for the incoming (towards the surface) 
and $t>0$ as the outgoing portions of the trajectory.
The PDOS in \eqref{eq:occu} is given in the $z$-representation as
\begin{equation}
	\frac{1}{v^{2}}|p(\ene,z)|^{2} = \left|\frac{1}{v}\int_{z_{0}}^{z} dz'\sqrt{\frac{\Delta(z')}{\pi}} \
	B(z')\exp\left\{-\frac{i}{v}\int_{z'}^{z} dz'' \left[\tilde{\ene}_{a}(z'') -i\Delta(z'') -\ene \right]\right\}\right|^{2}.
	\label{eq:pdos_z}
\end{equation}
Consequently, the adsorbate orbital occupancy can be straightforwardly expressed in the $z-$ representation
using \eqref{eq:pdos_z}.

\subsection{Energy dissipation}
\label{subsection:energy_trans}
 
A moving adsorbate needs to dissipate its kinetic energy to be adsorbed on the surface.
Possible dissipation channels under an electrochemical condition are the excitation 
of vibration in solution and the creation of e-h pairs in metal electrode. 
With the polaron transformed Hamiltonian \eqref{eq:lf_totham}, 
let us investigate the total amount of the energy dissipation by integrating the rate of change 
as derived in Appendix \ref{appendix:en_ex_rate}
\begin{equation}
	\dot{\mathcal{E}} = \langle \dot{\tilde{H}} \rangle  = \dot{\tilde{\ene}}_{a}(t)\langle n_{a}(t) \rangle \
	+ \frac{\dot{\Delta}(t)}{\sqrt{\pi\Delta(t)}}\int d\ene f(\ene) \Im[p(\ene,t)],
	\label{eq:en_tr}
\end{equation}
where $p(\ene,t)$ is given by \eqref{eq:p_et}. 
Note that \eqref{eq:en_tr} is the same as that derived for vacuum condition\cite{Mizielinski2005}
with the exception of the renormalized adsorbate energy level and the presence of solvent correlation 
function in $p(\ene,t)$.
On the other hand, the amount of average energy transferred non-adiabatically $\bar{E}$
can be obtained by integrating the rate below, which is derived following 
the substitution of $\langle n_{a}(t) \rangle$ 
with $\langle \delta n_{a}(t) \rangle = \langle n_{a}(t) \rangle - \langle n_{a}^{ad}(t) \rangle$
and $p(\ene,t)$ with $\delta p(\ene,t) = p(\ene,t) - p^{ad}(\ene,t)$
\begin{equation}
	\dot{\bar{E}} = \dot{\tilde{\ene}}_{a}(t)\langle \delta n_{a}(t) \rangle +\
	\frac{\dot{\Delta}(t)}{\sqrt{\pi\Delta(t)}}\int d\ene f(\ene) \Im[\delta p(\ene,t)].
	\label{eq:en_tr_na}
\end{equation}
Just as in the non-adiabatic orbital occupancy,
the non-adiabatic average energy transfer rate \eqref{eq:en_tr_na} can be expressed in $z$-representation by
changing the time derivatives to derivatives with respect to $z$. The resulting expression is 
valid for all constant values of $v$. By numerically integrating \eqref{eq:pdos_z}, both $\langle \delta n_{a}(z) \rangle$
and $\delta p(\ene,z)$ are easily obtained and hence $\dot{\bar{E}}$.

\subsubsection{Electronic friction coefficient}

The e-h excitations near the Fermi level of the electrode induced by
the approaching adsorbate give rise to a frictional force
that slows it down (see \figref{fig:system}). 
Within the trajectory approximation, the nuclear dynamics is essentially described by the Langevin
equation ($m=1$)
\begin{equation}
	\ddot{z}(t) = -\frac{dU(z)}{dz} - \int_{t_{0}}^{t}d\tau K(t-\tau)v(\tau) + \zeta(t),
	\label{eq:langevin}
\end{equation}
where $U(z)$ is the adsorbate-electrode adiabatic potential energy surface, $K(t-\tau)$ is 
the dissipative nonlocal kernel that depends on the memory of the system and $\zeta(t)$
is the random force. The second term in \eqref{eq:langevin} is the frictional force and is
related to $\zeta(t)$ through the fluctuation-dissipation theorem. Since we are interested in the
force induced by low energy excitations, we assume that the frictional force is purely electronic.
In the limit of slowly moving adsorbate (small $v(t)$), the friction kernel becomes local
(Markovian) i.e., $K(t-\tau)=\delta(t-\tau)\eta(t)$, where $\eta(z(t))$ is the  electronic friction coefficient.
We do not wish to evaluate the full dynamics of adsorbate using \eqref{eq:langevin}. Instead, our goal
is to derive an expression for the local electronic friction coefficient in the slow motion or equivalently,
in the nearly adiabatic limit. Evaluating the frictional force integral in Markovian limit gives
\begin{equation}
	\int_{t_{0}}^{t}d\tau K(t-\tau)v(\tau) = \eta[z(t)]v(t)\equiv\frac{\dot{\bar{E}}^{SM}}{v(t)},
	\label{eq:fric_force}
\end{equation}
where, 
\begin{equation}
	\dot{\bar{E}}^{SM} = \dot{\tilde{\ene}}_{a}(t)\langle \delta n_{a}^{SM}(t) \rangle +\
	\frac{\dot{\Delta}(t)}{\sqrt{\pi\Delta(t)}}\int d\ene f(\ene) \Im[\delta p^{SM}(\ene,t)].
	\label{eq:en_tr_slm}
\end{equation}
is the average energy transfer rate in the slow motion (SM) limit. For a vanishingly small values of 
$v$, $t$ becomes sufficiently large such that $B(t)\sim 1$. Following \cite{Mizielinski2005}, 
we proceed by performing Taylor expansions of the time-dependent quantities near their adiabatic values
\begin{equation}
	\begin{aligned}
		\tilde{\ene}_{a}(t') \approx \tilde{\ene}_{a}(t)+(t-t')\dot{\tilde{\ene}}_{a}(t) \\
		\Delta(t') \approx \Delta(t)+(t-t')\dot{\Delta}(t),
\end{aligned}
	\label{eq:taylor}
\end{equation}
and inserting them into \eqref{eq:p_et} resulting in 
\begin{equation}
	\begin{aligned}
	p(\ene,t) = & \sqrt{\frac{\Delta(t)}{\pi}} \int_{-\infty}^{0} d\tau  \
	\lsb 1 +i\frac{\dot{E}_{a}(t)}{2}\tau^{2} - \tau\frac{\dot{\Delta}(t)}{2\Delta(t)} \rsb \\
	\times & \exp\left\{-i\left[E_{a}(t) -\ene \right]\tau \right\}.
\end{aligned}
	\label{eq:pna}
\end{equation}
where $E_{a}(t) \equiv \tilde{\ene}_{a}(t) -i\Delta(t)$. To arrive at \eqref{eq:pna}, we expanded the term proportional
to the exponential of the small quantity $\dot{E}_{a}(t)$ 
and changed the integration to $\tau=t-t'$ with $t_{0}\rightarrow -\infty$.
Performing the Gaussian integration of the first term is straightforward and yields the 
adiabatic expression for $p$ and hence recover $\langle n_{ad} \rangle$ in \eqref{eq:nad}.
The succeeding terms in \eqref{eq:pna} correspond to the 
non-adiabatic corrections due to the adsorbate motion in SM limit.
The nearly adiabatic orbital occupancy 
then takes the form
\begin{equation}
	\begin{aligned}
		\langle n_{a}(t) \rangle  & = \int d\ene f(\ene) |p^{ad}(\ene,t) + \delta p^{SM}(\ene,t)|^{2} \\
		& =  \langle n_{a}^{ad}(t) \rangle +\langle \delta n_{a}^{SM}(t) \rangle,
	\end{aligned}
	\label{eq:occupation_na}
\end{equation}
where $\delta p^{SM}(\ene,t)$ is given by the 2nd and 3rd terms of \eqref{eq:pna},  
and the non-adiabatic component of the occupancy  $\langle \delta n_{a}^{SM}(t) \rangle$
in the SM limit may be expressed as
\begin{equation}
	\langle \delta n_{a}^{SM}(t) \rangle   = -\pi\int d\ene \frac{df(\ene)}{d\ene}\
	\left\{\frac{\dot{\Delta}(t)}{\Delta(t)}\left[\ene-\tilde{\ene}_{a}(t) \right] +\dot{\tilde{\ene}}_{a}(t)\right\}\
	\rho^{2}_{a}(\ene,t),
	\label{eq:occ_na1}
\end{equation}
from which we obtain the first term of \eqref{eq:en_tr_slm}. To this end, 
we discarded the negligible terms proportional to the
products $\delta p^{*SM}(\ene,t) \delta p^{SM}(\ene,t)$. We then performed integration by parts,
and ignored the neglible first term.
Here, $\rho_{a}(\ene,t)$ is the adsorbate PDOS.
Following similar procedure described above, second term of \eqref{eq:en_tr_slm} is given by 
\begin{equation}
	\begin{aligned}
	\frac{\dot{\Delta}(t)}{\sqrt{\pi\Delta(t)}}\int d\ene f(\ene) \Im[\delta p^{SM}(\ene,t)] &= \
	-\pi\int d\ene \frac{df(\ene)}{d\ene} \frac{\dot{\Delta}(t)}{\Delta(t)}\left[\ene-\tilde{\ene}_{a}(t)\right] \\
	&\times \left\{\frac{\dot{\Delta}(t)}{\Delta(t)}\left[\ene-\tilde{\ene}_{a}(t)\right] +\dot{\tilde{\ene}}_{a}(t)\right\}\
	\rho^{2}_{a}(\ene,t).
\end{aligned}
	\label{eq:en_tr_2t1}
\end{equation}
Substituting \eqref{eq:occ_na1} and \eqref{eq:en_tr_2t1} into \eqref{eq:en_tr_slm} yields
\begin{equation}
	\dot{\bar{E}}^{SM} =-\pi\int d\ene \frac{df(\ene)}{d\ene}\
	\left\{\frac{\dot{\Delta}(t)}{\Delta(t)}\left[\ene-\tilde{\ene}_{a}(t)\right] +\dot{\tilde{\ene}}_{a}(t)\right\}^{2}\
	\rho^{2}_{a}(\ene,t).
	\label{eq:en_tr_slm1}
\end{equation}
In the $z$-representation, and by comparing with \eqref{eq:fric_force},
$\dot{\bar{E}}^{SM}$ takes the form
\begin{equation}
	\dot{\bar{E}}^{SM}(z(t)) =\eta(z)v^{2},
	\label{eq:en_tr_slm2}
\end{equation}
where the coefficient of electronic friction is given by
\begin{equation}
	\eta(z) =-\pi\int d\ene \frac{df(\ene)}{d\ene}\
\left\{\frac{d\Delta(z)}{dz}\frac{\left[\ene-\tilde{\ene}_{a}(z)\right]}{\Delta(z)} +\frac{d\tilde{\ene}_{a}(z)}{dz}\right\}^{2}\
	\rho^{2}_{a}(\ene,z).
	\label{eq:friction}
\end{equation}
Apart from the presence of the renormalized adsorbate energy level, our above result
is equivalent to that of \cite{Mizielinski2005}. Further, by setting $\Delta$ independent of $z$, 
\eqref{eq:friction} reduces to the expression
of the electronic friction coefficient in vacuum\cite{Dou2018} where the adsorbate nuclear motion
has been considered. In most cases, the relevant electrons that participate in
excitation reside near $\ene_{F}$. We may take the surface temperature to be zero such that
the derivative of the Fermi-Dirac distribution becomes a delta function,
and the energy integration in \eqref{eq:friction} is trivial giving 
\begin{equation}
	\eta(z) =\pi\
	\left\{\frac{d\Delta(z)}{dz}\frac{\left[\ene_{F}-\tilde{\ene}_{a}(z)\right]}{\Delta(z)} +\frac{d\tilde{\ene}_{a}(z)}{dz}\right\}^{2}\
	\rho^{2}_{a}(\ene_{F},z).
	\label{eq:friction_zeroT}
\end{equation}

\subsubsection{Electron-hole excitation}

It is suggestive to investigate how exactly the electron-hole excitations
arise from the perspective of the electrons of the metal electrode using 
the Hamiltonian \eqref{eq:totham}. Working again in the slow-motion limit,
the approaching adsorbate is seen by the metal electrons as a slowly varying 
perturbation\cite{Brako1981JPC,Brako1980}, which results in instantaneous 
phase shifts at the Fermi level. This phase shift is related to the 
electronic friction and hence it is possible to derive \eqref{eq:friction_zeroT},
provided that the low energy excitations are in the neighborhood of $\ene_{F}$.
Borrowing the methods from surface science, the probability $P(\ene)$ of a system
being excited with an energy $\ene$ after the perturbation is switched off at
$t=\infty$ is\cite{Gunnarsson1980,Schornhammer1981,Brako1980,Brako1981JPC}
\begin{equation}
	P(\ene) = \frac{1}{2\pi}\int_{-\infty}^{\infty}dte^{i\ene t}P(t)
	\label{eq:Pex}
\end{equation}
where
\begin{equation}
	P(t) =\langle \psi(\infty) | \exp\left\{-i\left[ H(\infty) - E_{0} \right]t \right\} | \psi(\infty) \rangle.
	\label{eq:Pext}
\end{equation}
The electrons are assumed to be in their ground state $|\psi(0) \rangle$ at $t=0$ with an energy
$E_{0} = \langle \psi(0) | H_{0} |\psi(0) \rangle$ before the perturbation was switched on.
The excited state $|\psi(\infty) \rangle$ is obtained by applying the time evolution operator $U(\infty,0)$
to $|\psi(0) \rangle$. Doing so in \eqref{eq:Pext} and switching from Schr\"{o}dinger to Heisenberg picture,
with $H_{H}(\infty)=U(\infty,0)H(\infty)U(\infty,0)$, yields
\begin{equation}
	P(t) =\langle \psi(0) | \exp\left[ -iH_{H}(\infty) t \right]  | \psi(0) \rangle e^{iE_{0}t}.
	\label{eq:Pext1}
\end{equation}
To obtain $H_{H}(\infty)$, we need the expressions for the electron operators $\cc$ and $\ca$ in the Heisenberg
picture. The procedure is rather lengthy, and we only summarize the important steps and results here.
For more details, we refer the readers to the original paper of Brako and Newns in Reference \cite{Brako1981}.
To proceed, one begins by inserting 
expression \eqref{eq:at_insert2} for annihilation operator $\aan$ that was derived using equations
of motion into \eqref{eq:intcheom} to get $\ca$. 
The resulting expressions for $\ca$ is then
simplified in the slow limit of the motion. Doing similarly for $\cc$,
the electronic Hamiltonian of the metal electrode may be written as
\begin{equation}
	H_{H}(\infty)=\sum_{k}\ene_{k}\cc(\infty)\ca(\infty)=\frac{i}{2\pi\rho(\ene_{k})}\
	\sum_{k',k''}\int dt \int ds \delta(t-s)\dot{F}_{k',k''}(t,s)c^{\dagger}_{k''}c_{k'},
	\label{eq:Hamex}
\end{equation}
where $\dot{F}_{k',k''}(t,s)=\frac{\partial}{\partial t}\left\{e^{i\ene_{k''}s}e^{-i\ene_{k'}t}e^{-2i\left[\delta(\ene_{k'},t)-\delta(\ene_{k''},s) \right]} \right\}$,
and $\delta(\ene_{k},t)=\arctan\left[ \frac{\Delta(t)}{\ene_{k}-\tilde{\ene}_{a}(t)}\right]$ is the phase shift.
Integrating over $s$, \eqref{eq:Hamex} takes the form
\begin{equation}
	H_{H}(\infty)=H_{0} + \sum_{k,k'}U_{k,k'}\cc c_{k'},
	\label{eq:Hamex1}
\end{equation}
where $H_{0} = \sum_{k}\ene_{k}\cc\ca$ is the unperturbed Hamiltonian of the metal electrons and 
$U_{k,k'}=\frac{1}{\pi\rho(\ene_{k})}\int dt \dot{\delta}(\ene_{k},t)\exp\left[i(\ene_{k}-\ene_{k'})t \right]$
may be considered as an off-shell scattering matrix from state $k$ to $k'$. The energy difference between
the two states $\Omega=\ene_{k'}-\ene_{k}$ is the e-h excitation energy. $\Omega$ is typically small
on the order of the inverse of the time scale by which $\dot{\delta}$ varies. Surface electrons that are
excited typically reside near the Fermi level so that it is sufficient to replace $\ene_{k}$
by $\ene_{F}$ in $\dot{\delta}$. \eqref{eq:Hamex1}
provides two main ways to calculate $P(t)$: The first is through the linked cluster expansion as done by
Brako and Newns in \cite{Brako1981}, and the second is through Tomononaga bosonization as proposed by Schr\"{o}nhammer and 
Gunnarsson \cite{Gunnarsson1980}. Either method yields the same formally exact expression for $P(t)$ at finite temperature
as
\begin{equation}
	P(t)=\exp\left\{ \frac{1}{\pi^{2}}\int_{0}^{\infty}d\Omega \Omega\
	|\delta(\Omega)|^{2}\left[\coth\left(\frac{\beta \Omega}{2}\right)(\cos\Omega t -1) -i\sin\Omega t \right]  \right\},
	\label{eq:Pt1}
\end{equation}
where $\delta(\Omega)=\int e^{i\Omega t}\dot{\delta}(\ene_{F},t)$. 
This allows us to calculate $P(\ene)$ through \eqref{eq:Pex}.
The strength of the delta function of $P(\ene)$ corresponds to 
the Debye-Waller factor that describes the probability of the 
system to remain in its ground state. Computing for the 
second moment of $P(\ene)$ gives the average energy transfer
\begin{equation}
	\bar{\epsilon}(t)=\frac{1}{\pi^{2}}\int d\Omega \Omega^{2} \
	|\delta(\Omega)|^{2}=\frac{1}{\pi}\int_{-\infty}^{\infty}dt \dot{\delta}(\ene_{F},t)^{2}.
	\label{eq:en_tr_ehex}
\end{equation}
This allows us to write the average energy transfer rate as
\begin{equation}
	\dot{\bar{\epsilon}}(t)=\frac{1}{\pi}\dot{\delta}(\ene_{F},t)^{2}=\pi \
	\left\{\frac{\dot{\Delta}(t)}{\Delta(t)}\left[\ene_{F}-\tilde{\ene}_{a}(t)\right] +\dot{\tilde{\ene}}_{a}(t)\right\}^{2}\
	\rho^{2}_{a}(\ene_{F},t),
	\label{eq:en_tr_ehex}
\end{equation}
which is nothing but energy transfer rate in SM limit \eqref{eq:en_tr_slm} at $T=0$.
From \eqref{eq:en_tr_ehex}, the coefficient of electronic friction in \eqref{eq:friction_zeroT} is 
straightforwardly obtained.

\section{Application}
\label{section:Application}
To illustrate the non-adiabatic effects on electron transfer in solvated systems, 
we apply the above formalism to the case of electrochemical
proton ($Z=1$) discharge on a metal electrode. Such a reaction is described by
the well known Volmer process $\mathrm{H_{3}O^{+}} + e^{-}\rightarrow \mathrm{H^{*}}+\mathrm{H_{2}O}$
which involves the dissociation of $\mathrm{H^{+}}$ from $\mathrm{H_{3}O^{+}}$ followed by
deposition on the metal electrode\cite{Levich1970,Kuznetsov1989,Schmickler1996}.
This proton coupled electron transfer (PCET) reaction is 
often treated within the Born-Oppenheimer approximation and  assumed to be fully
adiabatic, i.e., large electronic overlap between the proton and the electrode.
As a consquence, $\mathrm{H^{+}}$ is almost always guaranteed to be adsorbed.
In reality, the motion of the proton affects the electron transfer rate
and its fate of being captured by the metal depends upon how it is able to efficiently transfer
its kinetic energy towards various energy dissipation channels.
We explore the consequences of non-adiabaticity in the following by presenting some 
illustrative numerical calculations.
For simplicity, we omit considerations of bond beaking in $\mathrm{H_{3}O^{+}}$
and focus solely on the dynamics of the already dissociated $\mathrm{H^{+}}$.
The effects of bond breaking were addressed using a position-dependent 
potential term in a modified version of the Newns-Anderson-Schmickler model\cite{Santos2006,Santos2008}.
Although one can derive an adiabatic trajectory using this potential term,
the essence of non-adiabatic electron transfer dynamics may be captured by starting
from a given trajectory whereby the simplest version \eqref{eq:totham} was derived.
In the present case, we ignore the quantum tunneling effects of proton during its transit. 
Inclusion of these effects introduces additional complexities that merit a separate discussion, 
which we plan to address in future works.
Additionally, we do not consider adsorption and assume that the proton only scatters upon interacting with the metal.
As mentioned earlier, the scattering trajectory is classically described by $R(t)\approx z(t)=v|t|$
where we disregard any motion parallel to the surface.
Here, $t<0$ and $t>0$ denote incoming (towards the surface) and outgoing (away from the surface)
portions of the trajectory, respectively. 
For convenience, we define the turning point (closest to the electrode) of the trajectory as $z\equiv0.0$ Bohr.
It is important to note that this does not imply direct contact with the surface of the electrode.
Considering a proton adsorbate, we assume an initial orbital occupancy $\langle n_{a}(t_{0}) \rangle = 0$
and an orbital energy level equal to the Fermi energy ($\ene_{F}$) which is shifted by $\lambda$ ($cf.$ \figref{fig:system}).  
The time-dependent parameters $\Delta[z(t)]$ and $\tilde{\ene}_{a}[z(t)]$ are assumed to be Gaussian functions (see \figref{fig:parameters})
\begin{equation}
	\begin{aligned}
		\Delta(z) &=  \Delta\exp\left(-\alpha z^{2}\right) \\
		\tilde{\ene}_{a}(z) &= (\ene_{0}-\ene_{\infty})\exp\left(-\kappa z^{2}\right) + \ene_{\infty} + \lambda
	\label{eq:params}
\end{aligned}
\end{equation}
where $\alpha=\kappa=0.015~\alpha_{0}^{-2}$ ($\alpha_{0}$ is the Bohr radius) is a parameter 
describing the decay of the electrode's wavefunction into the solvent.
$\Delta=0.01~$eV and $\ene_{0}=-1.5~$eV are the values of the resonance width and adsorbate energy level at 
$z=0.0~$Bohr, and $\ene_{\infty}=\ene_{F}=0.0~$eV is the value of energy level when $z=\infty$.
$\Delta(z)$ is maximum at $z=0.0$ Bohr and $0$ at $z=\pm\infty$.
The parametrization of $\ene(z)$ describes a scenario in which an initially empty adsorbate orbital level
crosses $\ene_{F}$ and becomes occupied through the electron transfer from the metal surface.
Further details on the numerical evaluation can be found in  Appendix\ref{appendix:numerical}.

\subsection{Time-dependent proton coupled electron transfer}

\begin{figure}
  \centering
  \mbox{
	  \subfloat[\label{a}]{\includegraphics[scale=0.32]{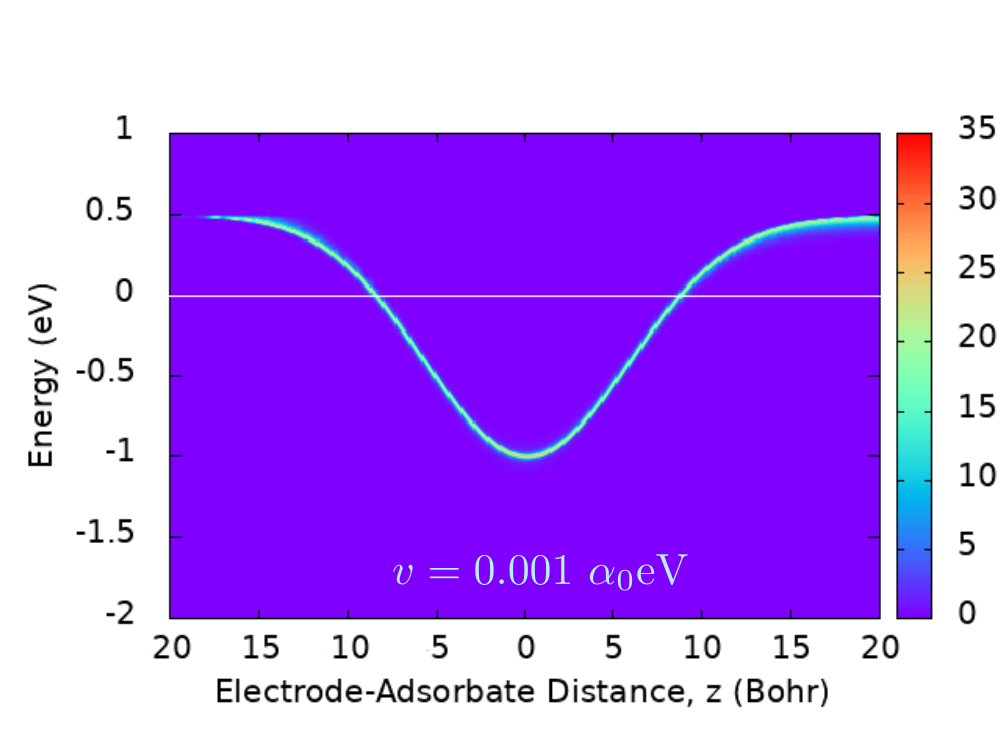}\label{fig:v0001}}\quad
	  \subfloat[\label{b}]{\includegraphics[scale=0.32]{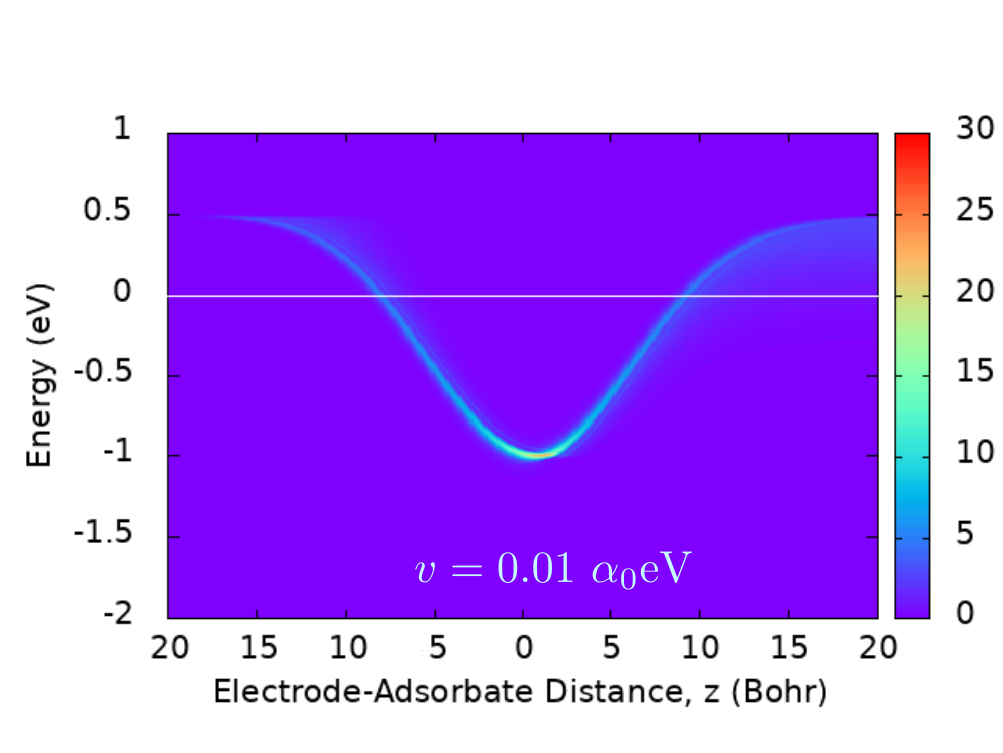}\label{fig:v001}}\quad
	  \subfloat[\label{c}]{\includegraphics[scale=0.32]{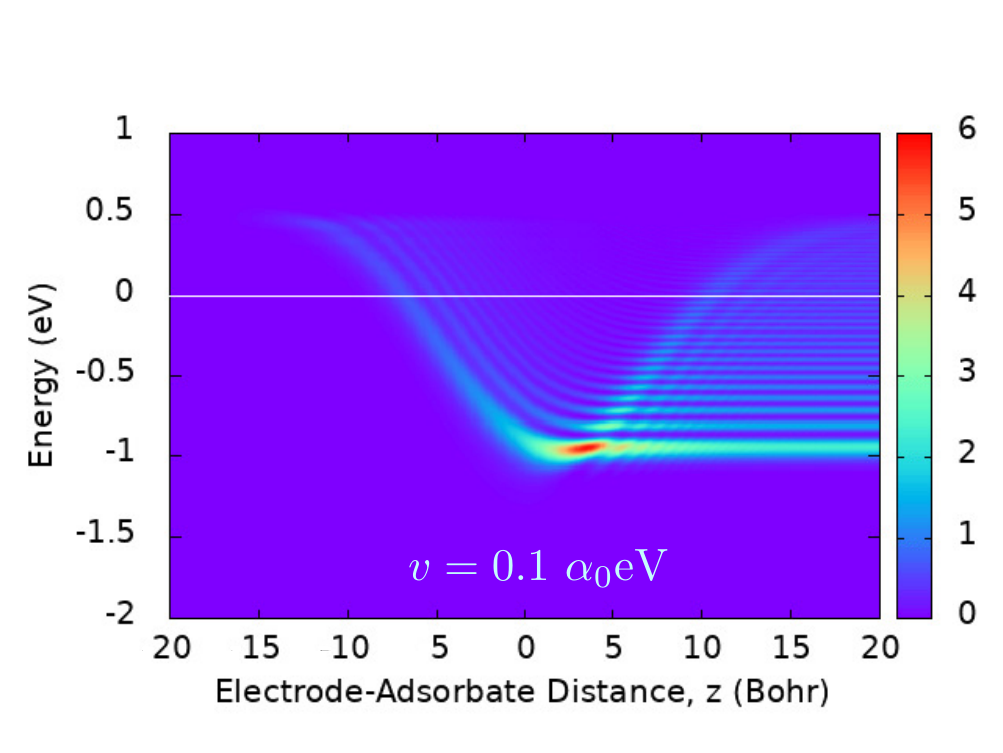}\label{fig:v01}}
  }
  \mbox{
	  \subfloat[\label{a}]{\includegraphics[scale=0.32]{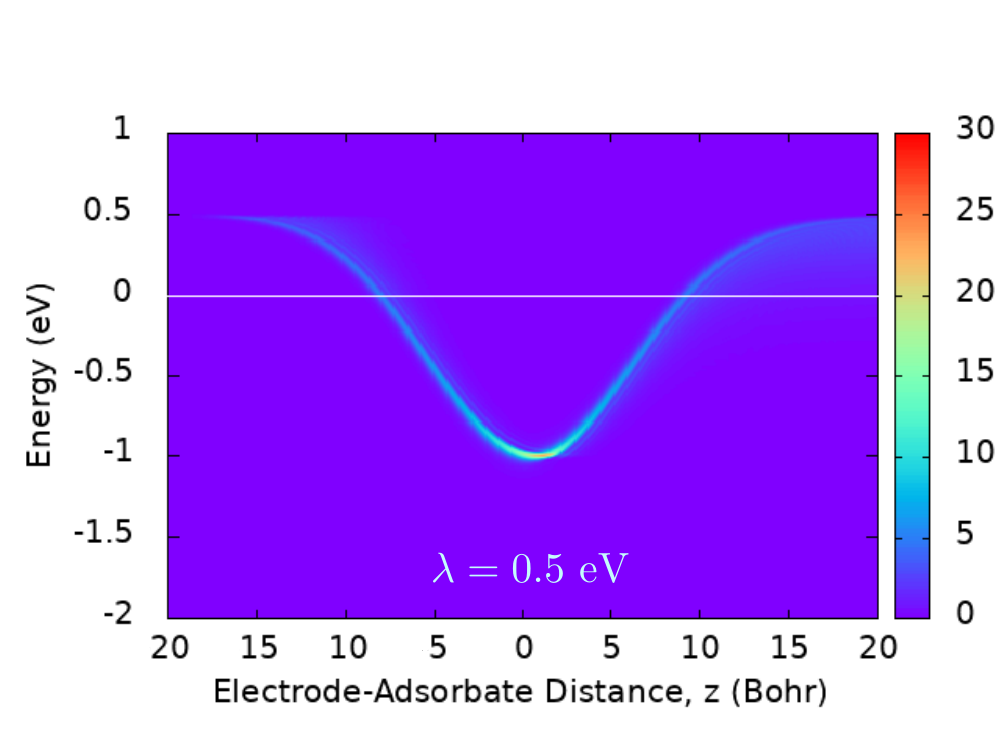}\label{fig:lamb05}}\quad
	  \subfloat[\label{b}]{\includegraphics[scale=0.32]{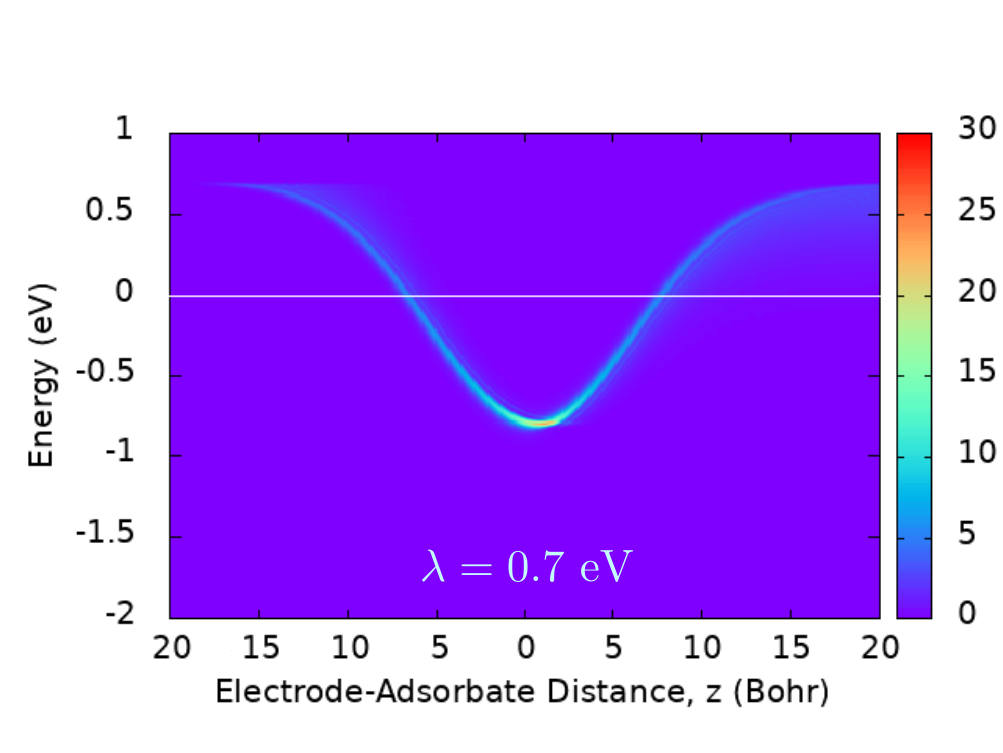}\label{fig:lamb07}}\quad
	  \subfloat[\label{c}]{\includegraphics[scale=0.32]{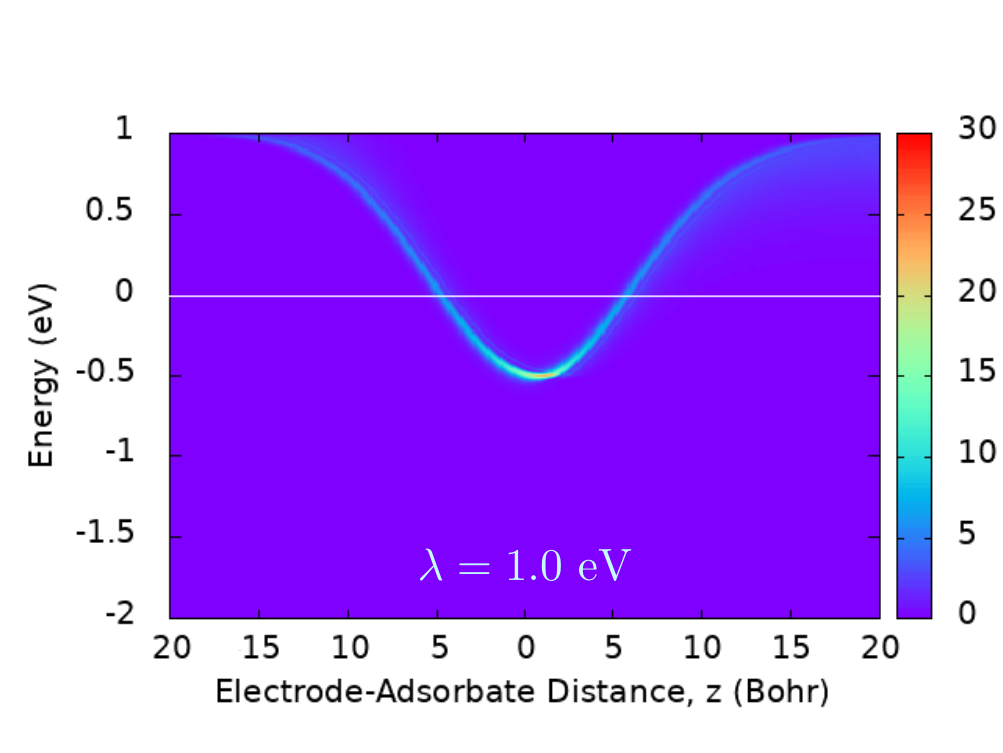}\label{fig:lamb1}}
  }
  \captionsetup{justification=raggedright,singlelinecheck=false}
  \caption{Time-dependent adsorbate projected density of states (PDOS) for different velocities (a)-(c) at 
  $\lambda=0.5~$eV and reorganization energies (d)-(e) at $v=0.01~\alpha_{0}$eV as functions of energy and adsorbate-electrode distance.
  The horizontal lines correspond to $\ene_{F}$. 
  Other parameters are $\Delta = 0.01$ eV, $\ene_{0} = -1.5$ eV, $\ene_{\infty} = \ene_{F}= 0.0$ eV,
  $\alpha = \kappa = 0.015~\alpha_{0}^{-2}$, and $T=300$ K. The left- and right-hand sides of $z = 0.0$ Bohr correspond to the
  incoming and outgoing parts of the trajectory, respectively.}
  \label{fig:2dpdos}
\end{figure}

\begin{figure}
\centering
\captionsetup{justification=raggedright,singlelinecheck=false}
\subfloat[\centering]{
	\includegraphics[scale=0.7]{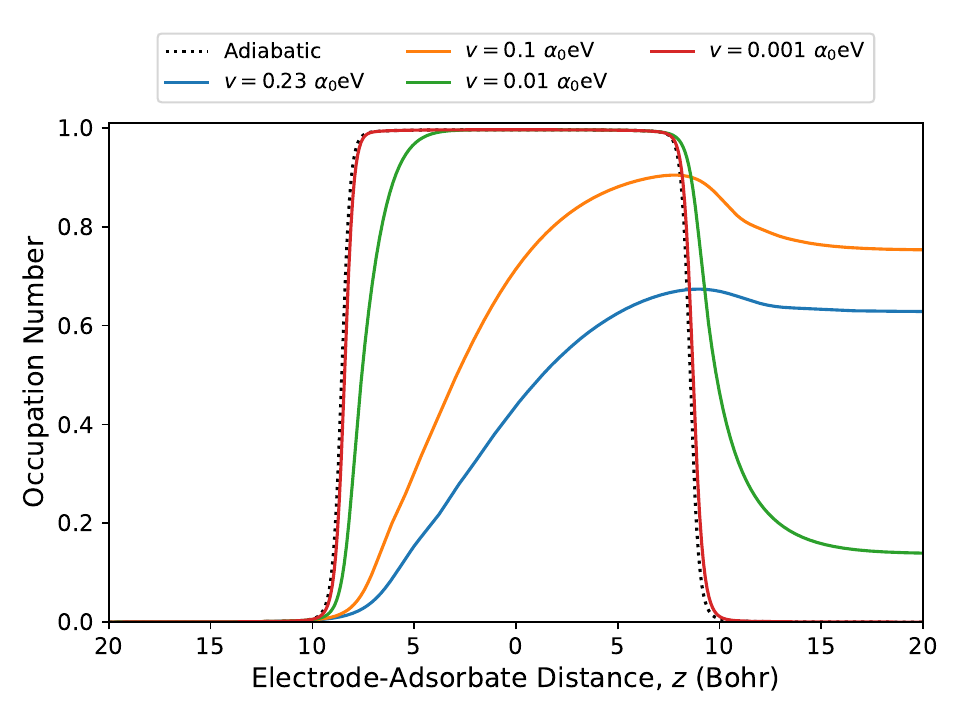} \label{fig:velvar}
}
\qquad
\subfloat[\centering]{
	\includegraphics[scale=0.7]{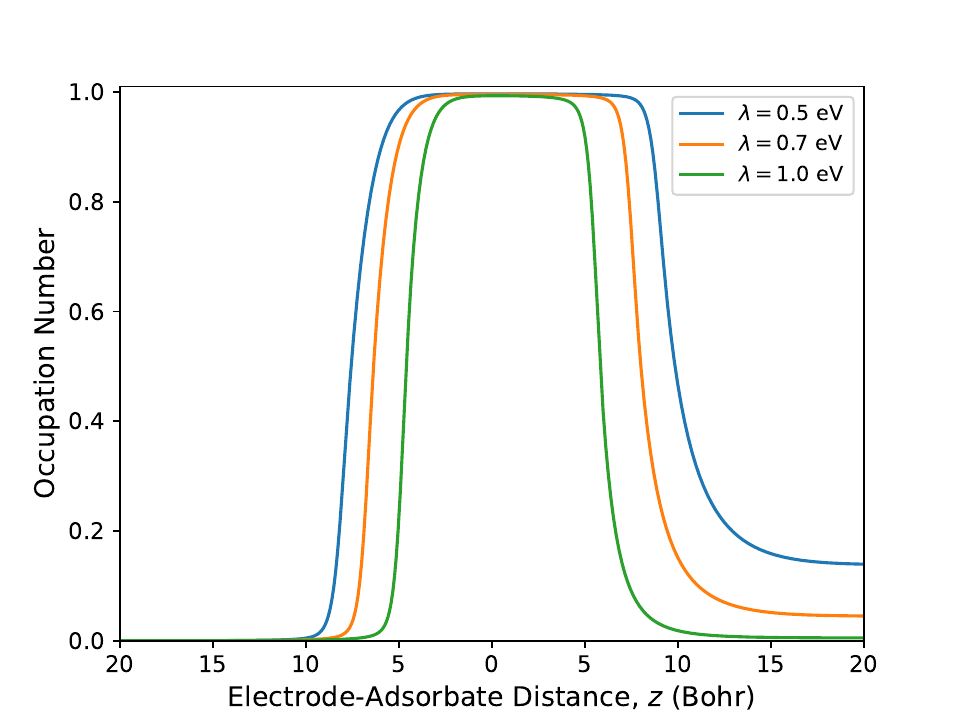} \label{fig:lambvar}
}
\caption{\label{fig:occu} Adsorbate orbital occupancy at different 
	(a) velocities when $\lambda=0.5$ eV and (b) reorganization energies when $v=0.01~\alpha_{0}$eV,
	using the parameters in \figref{fig:2dpdos}. The black dotted curves in (a) correspond
	to the adiabatic orbital occupancy.
}
\end{figure}

In \figref{fig:2dpdos} the plots illustrate the time-dependent PDOS (\eqref{eq:pdos_z}) obtained at $T=300$ K. 
The left- and right-hand sides of $z=0.0$ Bohr
represent the incoming and outgoing portions of the trajectory, respectively. 
The upper panel demonstrates the effects of 
velocity; (a) $v=0.001~\alpha_{0}$eV, (b) $v=0.01~\alpha_{0}$eV, and
(c) $v=0.1~\alpha_{0}$eV when $\lambda = 0.5$ eV. 
It is notable that the intensities are higher for smaller $v$ in accordance with \eqref{eq:pdos_z}.
The intensity corresponds to the matching of the energy $\ene$ with $\tilde{\ene}_{a}(z)$ (see \eqref{eq:pdos_z} or equivalently \eqref{eq:de}),
while the matching condition is relaxed because of $i\Delta$. The matching occurs once in the incoming and once in the outgoing
portions in addition to the case when $z=0.0$ Bohr.
For a slowed adsorbate ($v=0.001~\alpha_{0}$eV) the PDOS intensities are predominantly localized on a 
line given by $\ene=\tilde{\ene}_{a}(z)$.
As the adsorbate accelerates (\figref{fig:v01}), these intensities become asymmetrically broadened. 
The asymmetry arises from the retardation of the function $p(\ene,z)$ when $v$ is large.
In the lower panel, the effects of the reorganization energy are depicted:
(d) $\lambda=0.5$ eV, (e) $\lambda=0.7$ eV,and (f) $\lambda=1.0$ eV when $v=0.01~\alpha_{0}$eV.
The shape of the peak intensities is generally similar to \figref{fig:v001} while the intensity is shifted upward by
$\lambda$ as expected from the expression of $\tilde{\ene}_{a}(z)$, which is the main effect of e-ph coupling.
These findings corroborate the behavior of the adsorbate level occupations 
obtained by integrating the PDOS and the Fermi distribution functions over $\ene$ and are shown in \figref{fig:occu}. 

\figref{fig:velvar} shows the numerically calculated orbital occupancy of proton
for different velocities when $\lambda=0.5$ eV. In the figure, $v=0.23~\alpha_{0}$eV correspond to the 
thermal velocity.
The adiabatic occupation $\langle n_{a}^{ad}(z) \rangle$ (dotted lines) is highly symmetric in both incoming and outgoing portions 
of the trajectory with charge transfer occuring around $z\approx \pm 8.0$ Bohr.
Near $z=0.0$ Bohr, $\langle n_{a}^{ad}(z) \rangle = 1$ as expected since $\Delta$ is small. 
For non-zero $v$, the non-adiabatic effects manifest 
as significant deviations from $\langle n_{a}^{ad}(z) \rangle$
in both incoming and outgoing portions of the
trajectory. This is especially prominent at altitudes where $\tilde{\ene}_{a}$ crosses $\ene_{F}$.
For faster speeds, the non-adiabatic occupations $\langle n_{a}(z) \rangle$ are strongly reduced and 
are extremely asymmetric, with orbital fillings starting to occur closer to $z=0.0$ Bohr in the incoming trajectory.
When $v\ge 0.1~\alpha_{0}$eV, the orbital occupancies never reach 1 and are peaked at the outgoing portion.
Furthermore, noticeable fractional charge fillings persist long after the proton has left the scattering turning point 
($\tilde{\ene}_{a}(z)\le \ene_{F}$).
The reduction and asymmetry of $\langle n_{a}(z) \rangle$ can be easily deduced from the time-dependent
PDOS with retardation effects as shown in \figref{fig:2dpdos}.
Physically, these results highlight the inability of the electron system to return to its adiabatic
ground state when the adsorbate is in motion; the breakdown of the Born-Oppenheimer
approximation.
The electronic system relaxation time $\tau\propto 1/\Delta$ becomes large when $\Delta$ is small.
In electrochemical systems, $\Delta$ is typically small, on the order of a few meV
implying a much longer $\tau$, and enhanced non-adiabaticity.
Performing calculations with increased $\Delta$ when $v=0.1~\alpha_{0}$eV
results in the reduction 
of asymmetric electron transfer. For the case of adsorption in vacuum where $\Delta$ usually takes a large
value, a slight asymmetry in $\langle n_{a}(z) \rangle$ is also evident and enhanced by
increasing kinetic energy and metal work function\cite{Nakanishi1988}.
Additional non-adiabatic effects manisfest as overfillings of orbital occupancy\cite{Yoshimori1986}.
For much slower speeds, $\langle n_{a} (z)\rangle$ approaches the value of 
$\langle n_{a}^{ad}(z) \rangle$.
When $v=0.001~\alpha_{0}$eV for example, the non-adiabatic effects are suppressed and the orbital
occupancy very nearly coincides with the adiabatic value. This can also be seen in \figref{fig:v0001}
where PDOS retardations are vanishing and the peaks are highly localized. This confirms that our calculations
reduce to the adiabatic value in the limit of vanishing velocity.

The effects of coupling with the solvent environment are shown in \figref{fig:lambvar} for 
proton speed of $v=0.01~\alpha_{0}$eV. Within our chosen parameters, $\langle n_{a}(z) \rangle$
is visibly smaller and narrower when $\lambda=1.0$ eV, which
can be traced back to the behavior of the renormalized energy level depicted in
\figref{fig:en_level} in Appendix \ref{appendix:numerical} and the time-dependent PDOS with retardation in \figref{fig:2dpdos}. 
At this speed, several delocalized intensities are visible above $\ene_{F}$ (\figref{fig:v001}).
When $\lambda$ is increased, these are shifted upwards and do not contribute during
$\ene$ integration due to the energy range of $f(\ene)$, resulting in narrower orbital occupancy.
In contrast, for weaker $\lambda$, $\langle n_{a}(z) \rangle$ are broader
due to the Fermi level crossings occuring farther away from $z = 0.0$ Bohr.
The persistent partial charge occupations 
in the outgoing portion are due to the minimal shifting (compared to strong $\lambda$) of the 
PDOS intensities which enables the retardation effects to contribute during $\ene$ integrations.
When $v=0.1~\alpha_{0}$eV, the non-adiabatic effects are amplified and the main
effects of $\lambda$ is to reduce the occupation.

We obtain the distance-dependent electron transfer (reduction) rate by dividing the occupation 
with the electron system's relaxation time $1/2\Delta$, giving 
$k(z)=2\Delta(z)\langle n_{a}(z) \rangle$. The total rate $k$ is the sum of $k(z)$ 
over the whole path of the adsorbate. Employing similar procedure derives the adiabatic (static) rate $k_{ad}$
from \eqref{eq:nad}.
In Appendix \ref{appendix:marcus}, we show that in the $k_{B}T\gg \Delta$ limit, $k_{ad}$ reduces to the 
electron transfer rate in the Marcus theory. 
\begin{figure}
\centering
\captionsetup{justification=raggedright,singlelinecheck=false}
\subfloat[\centering]{
	\includegraphics[scale=0.7]{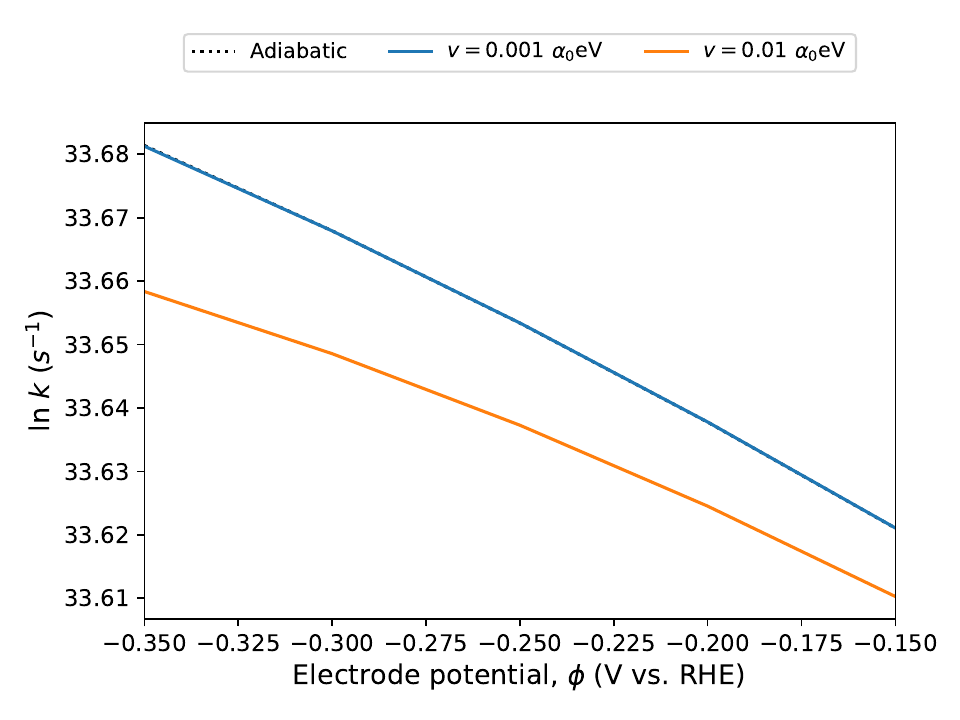} \label{fig:ratevelvar}
	\begin{picture}(0,0)
		\put(-125,125){\includegraphics[height=2.5cm]{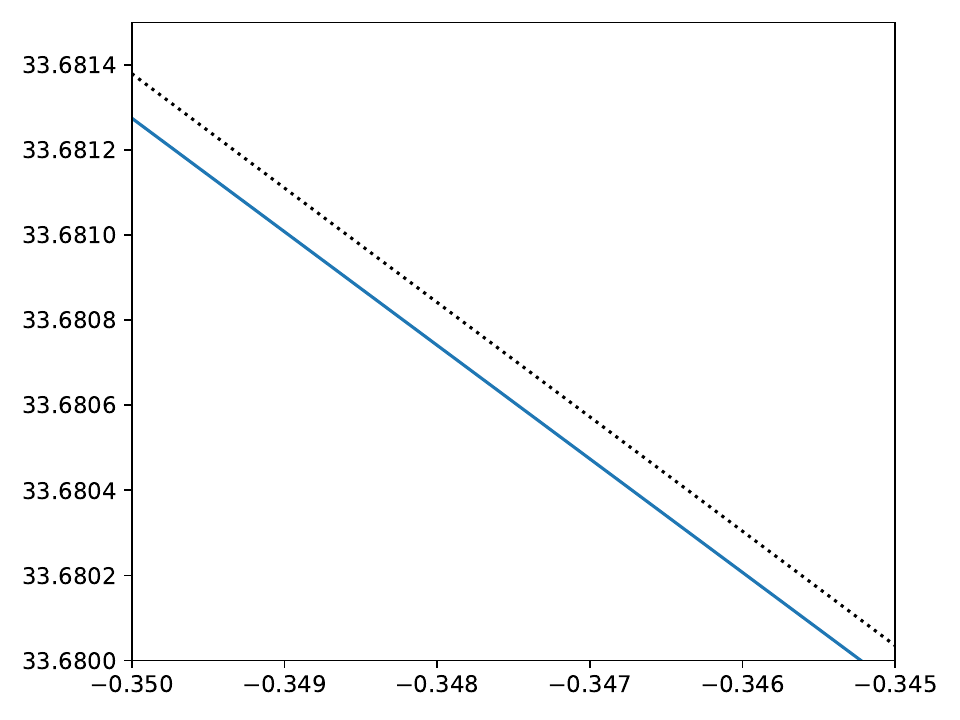}}
	\end{picture}
}
\qquad
\subfloat[\centering]{
	\includegraphics[scale=0.7]{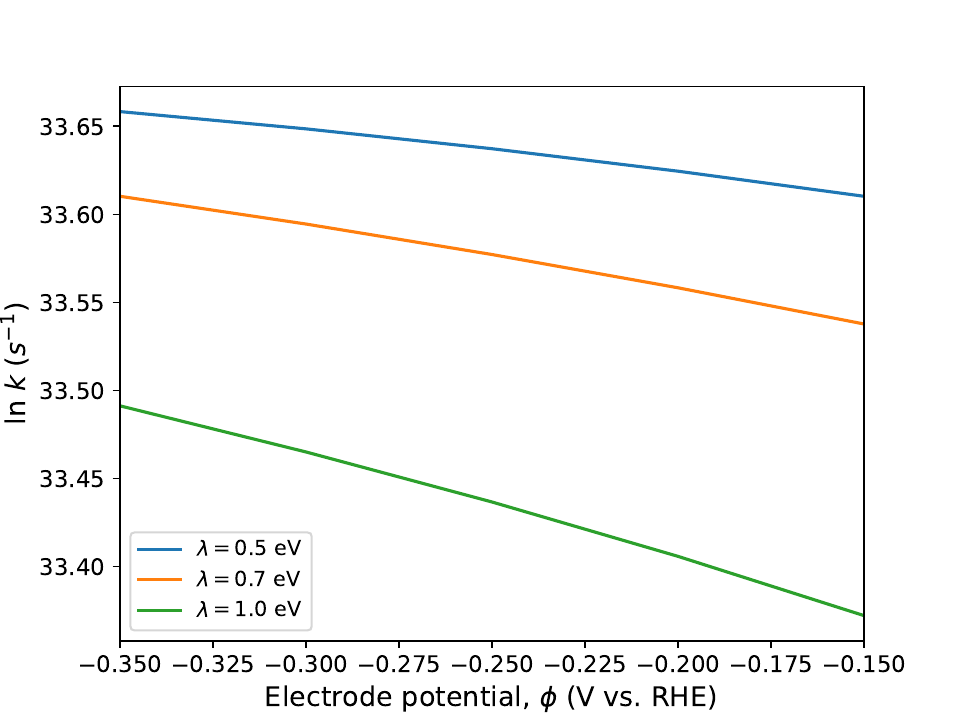} \label{fig:ratelambvar}
}
\caption{\label{fig:rate} Electron transfer rate for different 
	(a) velocities when $\lambda=0.5$ eV and (b) reorganization energy when $v=0.01~\alpha_{0}$eV
	as a function of electrode potential $\phi$ in Volts (V) vs. reversible hydrogen electrode (RHE) 
	using the same parameters as in \figref{fig:2dpdos}.
	 The black dotted line in (a) corresponds
	 to the adiabatic rate and the inset shows the zoomed rate for $-0.350 \leq \phi \leq -0.345$ V vs. RHE
}
\end{figure}
\figref{fig:rate} shows the electron transfer rates for different $v$ and $\lambda$ 
as functions of the electrode potential $\phi$ vs. reversible hydrogen electrode (RHE). 
The range of $\phi$ is chosen to conform with
the potential window of experimentally observed reduction rate of proton\cite{Hamelin1987}. 
The cathodic transfer coefficient $\alpha$ which is proportional to the Tafel slope
is observed to decrease (increase) at increasingly (decreasingly) negative $\phi$.
While we do not aim a quantitative comparison, to within our choice of parameters, we notice
that the slopes of our computed $k$ and hence $\alpha$ in both most and least negative regions of
$\phi$ are generally shown  to exhibit the same tendencies as the experimental observation.
\figref{fig:ratevelvar} shows that when the velocity is decreased, the electron transfer rate
$k$ approaches the adiabatic result for relatively weak coupling with solvent modes ($\lambda = 0.5$ eV). 
Similar conclusions can be drawn for other values of $\lambda$.
We find that $v=0.001~\alpha_{0}$eV is slow
enough to closely coincide with $k_{ad}$. At this speed, the system can be considered
as nearly adiabatic. In contrast, higher velocities significantly hinder the electron
exchange resulting in the decrease of $k$. This is especially prominent at thermal velocity
$v=0.23~\alpha_{0}$eV (not shown)
where $k$ is about four times lesser compared to $k_{ad}$. The effects of coupling with the solvent modes
shown in \figref{fig:ratelambvar} follow closely the results for $\langle n_{a}(z) \rangle$ 
at different strengths of $\lambda$, i.e., weak (strong) $\lambda$ results in higher (lower)
electron transfer rates. The solvent effectively screens the tunneling electron and renormalizes
the adsorbate energy level, resulting in the decrease of orbital occupancy and hence $k$.

\subsection{Energy transfer}

\begin{figure}
\centering
\captionsetup{justification=raggedright,singlelinecheck=false}
\includegraphics[scale=0.7]{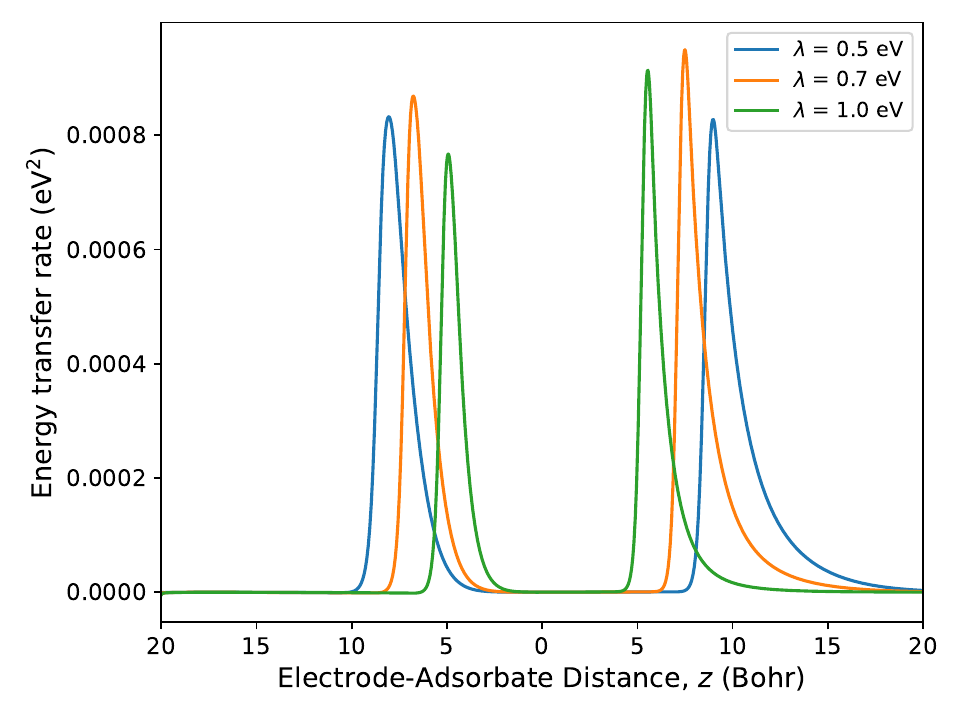}
\caption{\label{fig:en_trans} Average energy transfer rate 
of an adsorbate scattering on a metal electrode at speed of $v=0.01~\alpha_{0}$eV for
different strengths of coupling with solvent modes $\lambda$.
Other parameters are $\Delta = 0.01$ eV, $\ene_{0} = -1.5$ eV, $\ene_{\infty} = \ene_{F}= 0.0$ eV,
	and $\alpha = \kappa = 0.015~\alpha_{0}^{-2}$.
}
\end{figure}

Using the same parameters as in preceeding subsection, we numerically evaluate 
the non-adiabatic average energy transfer rate \eqref{eq:en_tr_na} 
for $v=0.01~\alpha_{0}$eV and different values of $\lambda$. The results are 
shown in \figref{fig:en_trans}.
For all strengths of $\lambda$ and in both incoming and outgoing parts of the trajectory, 
the $\dot{\bar{E}}(z)$ is peaked at the positions where $\tilde{\ene}_{a}(z)\simeq\ene_{F}$.
The peak position is closer to the turning point ($z=0.0$ Bohr) for stronger $\lambda$, in
accordance with the position dependent parameters discussed in Appendix \ref{appendix:numerical}.
It was shown\cite{Gunnarsson1980} previously that the largest contribution to the average energy transfer 
$\bar{E}$ is proportional to
$\sim 1/\Delta$.
This implies that $\bar{E}$ is
large when the Fermi level crossings occur at distances far from the metal surface, where $\Delta$ is small.
In our case, $\bar{E}$ can be obtained by integrating \eqref{eq:en_tr_na} over the 
whole path which yields the following tendency: 
$\bar{E}(\lambda=0.5~\mathrm{eV})>\bar{E}(\lambda=0.7~\mathrm{eV})>\bar{E}(\lambda=1.0~\mathrm{eV})$,
i.e., the proton loses more of its kinetic energy when the coupling with the solvent
modes is weak.
Clearly, from \figref{fig:en_trans}, Fermi level crossing occurs farthest from the metal electrode when $\lambda=0.5~$eV, 
supporting the above average energy loss tendency.
Within the trajectory approximation, the effects of velocity to the 
average energy exchange rate can be indirectly inferred from \figref{fig:occu} and 
the fact that $\dot{\bar{E}}$ is proportional 
to $\langle \delta n_{a} \rangle = \langle n_{a} \rangle - \langle n_{a}^{ad} \rangle$.
When the adsorbate velocity is small, $\dot{\bar{E}}$ features nearly symmetric peaks
centered at distances where Fermi level crossings occur. On the other hand, when
the adsorbate moves faster, we notice peculiar energy gains ($\dot{\bar{E}} < 0$)
on the way out of the scattering region. This is mainly due to $\langle \delta n_{a} \rangle$
being negative within this range as can be deduced from the difference
$\langle n_{a} \rangle - \langle n_{a}^{ad} \rangle$ in \figref{fig:occu}.

\begin{figure}
\centering
\captionsetup{justification=raggedright,singlelinecheck=false}
\subfloat[\centering]{
	\includegraphics[scale=0.7]{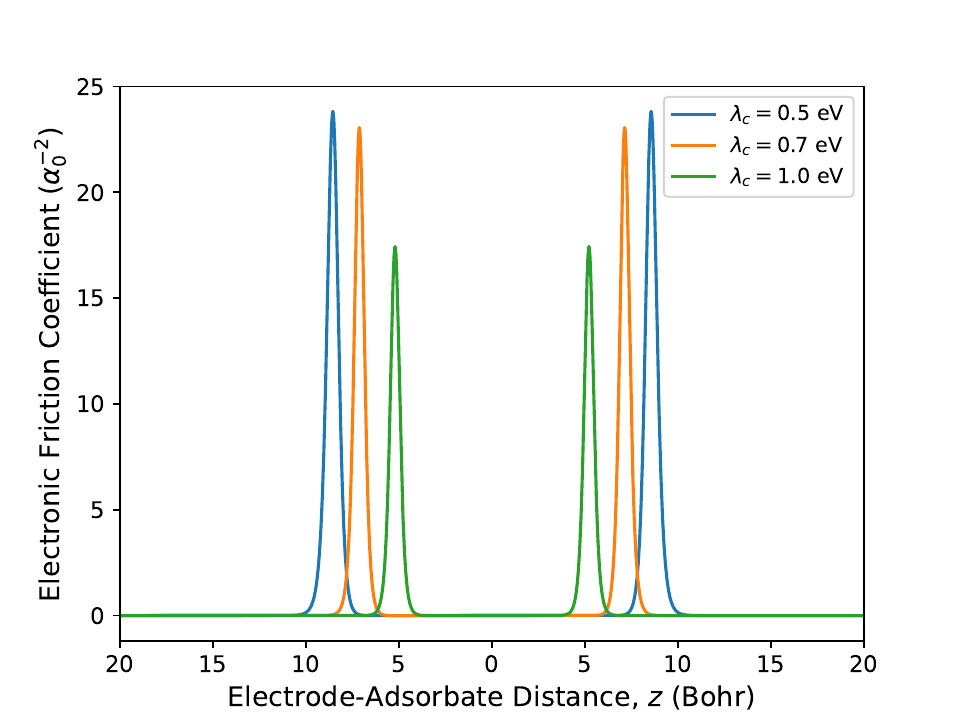} \label{fig:friclambvar}
}
\qquad
\subfloat[\centering]{
	\includegraphics[scale=0.7]{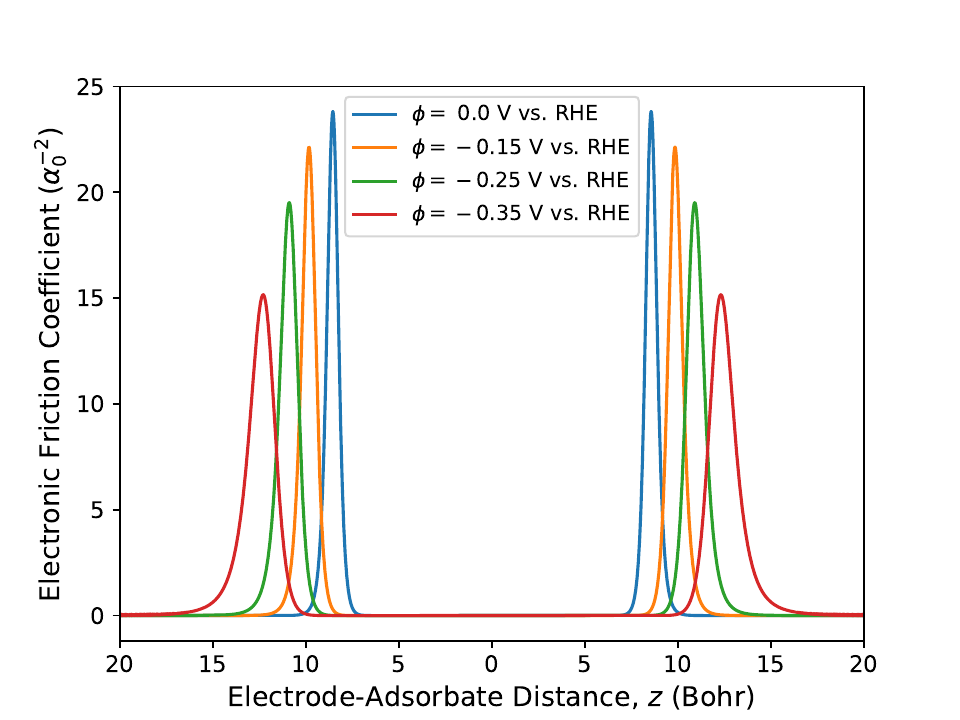} \label{fig:fricbiasvar}
}
\caption{\label{fig:friction} Electronic friction coefficient for different 
	(a) $\lambda$ when electrode potential $\phi=0.0$ V and (b) $\phi$ when $\lambda= 0.5$ eV
	as a function of $z$. Other parameters are the same as in \figref{fig:en_trans}. 
}
\end{figure}

In the slow motion limit, the average energy transfer rate is essentially dictated
by the electronic friction coefficient $\eta(z)$ (\eqref{eq:friction}).
\figref{fig:friclambvar} shows $\eta(z)$ at $T=300$ K for different values of the reorganization 
energy. The curves are highly symmetric in both portions
of the trajectory and exhibit similar tendencies with the average energy transfer rate
for non-negligible $v$ as seen in \figref{fig:en_trans}. The peaks are centered at an adsorbate 
altitude where the adsorbate energy level crosses the Fermi level. 
For weak $\lambda$, this level crossing occurs at
a distance far away from the metal implying significant energy loss toward e-h excitations.
The opposite tendency can be deduced when $\lambda$ is strong in which the peaks are shifted 
closer to the metal. This can be summarized by integrating $\dot{\bar{E}}^{SM}$ over the whole trajectory
to yield the average energy transfer (loss) in the SM limit which results in
$\bar{E}^{SM}(\lambda=0.5~\mathrm{eV})>\bar{E}^{SM}(\lambda=0.7~\mathrm{eV})>\bar{E}^{SM}(\lambda=1.0~\mathrm{eV})$.

The effects of the electrode potential on the electronic friction coefficient when $\lambda = 0.5$ eV is 
depicted in \figref{fig:fricbiasvar}. 
Negative values of the electrode potential $\phi$ means shifting
$\tilde{\ene}_{a}(z)$ downward relative to $\ene_{F}$,
and then shifting the Fermi level crossings away from the metal electrode.
When $\phi = -0.35$ V vs. RHE, for example, $\eta(z)$ is significantly broadened
and lowered.
This indicates that an increasing $\phi$ would make the frictional
force act on a much wider range of $z$, thus increasing the average
energy transfer.
This leads to the important consequence that increasing $\phi$ promotes the energy loss of
an adsorbate and increase the likelihood of trapping on the metal surface.

We further remark on the possibility of the proton being captured by
the metal surface during their encounter. Assuming that the relevant energy dissipation channel
is the e-h excitations in the metal, the sticking probability of the proton can be calculated
as
\begin{equation}
	S(E_{i})=\int_{E_{i}}^{\infty}P(\ene)d\ene,
	\label{eq:sticking}
\end{equation}
where $E_{i}$ is the initial kinetic energy of the proton, and $P(\ene)$
is the e-h excitation probability \eqref{eq:Pex}.
Roughly speaking, when the average energy loss is 
larger than $E_{i}$, $S \rightarrow 1$, or the proton is likely to be adsorbed.
Without resorting to the full numerical evaluation, 
it is possible 
obtain an analytical form of $P(\ene)$
at least in the high temperature 
limit. To the leading order, we may introduce approximations as  $\coth(\beta\Omega/2)\approx 2/\beta\Omega$,
$\cos\Omega t -1 \approx \Omega^{2}t^{2}/2$ and $\sin\Omega t\approx \Omega t$ when $\Omega$ is substantially small.
Using these, \eqref{eq:Pt1} can be expressed as
\begin{equation}
	P(t)\approx\exp\left(-\bar{\epsilon}k_{B}Tt -i\bar{\epsilon}t \right),
	\label{eq:Pt2}
\end{equation}
which after evaluating the time integrals in \eqref{eq:Pex} yields
a Gaussian form of $P(\ene)$ as
\begin{equation}
P(\ene)=\frac{1}{\sqrt{2\pi\bar{\epsilon}k_{B}T}}\exp\left[ -\frac{(\ene-\bar{\epsilon})^{2}}{4\bar{\epsilon}k_{B}T}\right],
	\label{eq:Pex1}
\end{equation}
which is peaked at the values of the average energy transfer rate $\bar{\epsilon}$. If we take
$\bar{\epsilon}=\bar{E}^{SM}$ since they are formally equivalent via \eqref{eq:friction_zeroT}, our estimated
tendency for $\bar{E}^{SM}$ suggests that the 
$P(\ene)$ spectra will be most shifted towards the energy loss portion ($\ene > 0$) for small $\lambda$ and more negative $\phi$.
Evaluating \eqref{eq:sticking},
using \eqref{eq:Pex1} and \eqref{eq:en_tr_ehex} for $v=0.001\alpha_{0}$eV gives 
$S(\lambda=0.5~\mathrm{eV}) > S(\lambda=0.7~\mathrm{eV}) > S(\lambda=1.0~\mathrm{eV})$.
This implies that strong $\lambda$ results in small energy loss and the adsorbate is 
less likely to be trapped. Conversely, weak $\lambda$ promotes energy loss and the probability
of the adsorbate to stick on the metal surface.
We may also infer the effects of electrode potential on the e-h excitation
probability and the adsorbate sticking probability. For some values of $\lambda$, an increasingly negative
$\phi$ would shift the peaks of $P(\ene)$ towards the loss portion of the spectra. This indicates that
the sticking probability of the adsorbate is promoted when $\phi$ is increased.

\section{Summary and Conclusions}
\label{section:summary}

We have studied the non-adiabatic effects due to the dynamic adsorbate interacting 
with the metal electrode in the presence of solvent using the 
time-dependent Newns-Anderson-Schmickler model Hamiltonian. 
The electron transfer rate is studied in terms of the time-dependent 
adsorbate occupation derived using the nonequilibrium Green’s function
formalism and the trajectory approximation.
Our numerical calculations show that the electron transfer rate 
is significantly reduced for nonzero adsorbate velocities and 
large electron-bath phonons couplings. 
It is also shown that the rate is strikingly peaked 
at distances where the adsorbate energy crosses the Fermi level of metal electrode. 
These crossings occur away from (close to) the metal surface 
for the small (large) solvent reorganization energy $\lambda$.
This results in significant energy loss for weak $\lambda$.
For small adsorbate velocities, we derived the analytic expression of 
the electronic friction coefficient $\eta(z)$. 
$\eta(z)$ is strongly peaked at a distance where Fermi level crossings occur.
When the electrode potential is made more negative, the crossing occurs away 
from the metal surface leading to higher energy transfer rate and hence larger energy loss.
We also discussed the probability $P(\ene)$ of the electron system being excited,
from which we derived the average energy transfer rate and the coefficient of electronic friction.
In the high temperature limit, we discussed the sticking probability obtained from $P(\ene)$. 
Since $P(\ene)$ is peaked at the average energy loss value, 
we conclude that the sticking probability is high when $\lambda$ is small and the electrode potential is high.
\section{Acknowlegments}
This research is supported by the New Energy and Industrial Technology Development Organization (NEDO) project, 
MEXT as “Program for Promoting Researches on the Supercomputer Fugaku” 
(Fugaku battery \& Fuel Cell Project) (Grant No. JPMXP1020200301, Project No.: hp220177, hp210173, hp200131), 
Digital Transformation Initiative for Green Energy Materials (DX-GEM)
and JSPS Grants-in-Aid for Scientific Research (Young Scientists) No. 19K15397. 
Some calculations were done using the supercomputing facilities of the Institute for Solid State Physics,
The University of Tokyo.
\appendix
\section{Occupation Number from Equations of Motion Approach}
\label{appendix:eom}
We can recover the results of nonequilibrium Green's function formalism
using equations of motion approach (EOM).
The Heisenberg EOM for an operator $\hat{O}$ is given by
\begin{equation}
	\frac{d \hat{O}}{dt} = i [\tilde{H},\hat{O}] 
	\label{eq:heomop}
\end{equation}
where $\tilde{H}$ is the canonically transformed Hamiltonian.
For brevity, we drop the hats on the operators. The EOM of the adsorbate electron annihilation
operator is
\begin{equation}
	i\frac{d \tilde{a}(t)}{dt} =  \tilde{\ene}_{a}(t)\tilde{a}(t) + \sum_{k}V_{ak}(t)\ca(t),
	\label{eq:aheom}
\end{equation}
with $\tilde{V}_{ak}(t)\ac(t)=V_{ak}(t)\tilde{a}^{\dagger}(t)$.
We also require the EOM of metal electron's annihilation operator, 
\begin{equation}
		i\frac{d \ca(t)}{dt} =    \ek\ca(t) +V_{ak}^{*}(t)\tilde{a}(t)
	\label{eq:cheom}
\end{equation}
Integrating \eqref{eq:cheom} to solve for $\ca$ gives
\begin{equation}
	\ca(t) =  \exp\lsb -i\ek(t-t_{0}) \rsb \ca(t_{0}) -i \int_{t_{0}}^{t} dt' \exp\lsb -i\ek(t-t') \rsb V_{ak}^{*}(t')\tilde{a}(t'),
	\label{eq:intcheom}
\end{equation}
where $t_{0}$ is an initial reference time. We substitute this into \eqref{eq:aheom} to arrive at
\begin{equation}
	\begin{aligned}
		i\frac{d \tilde{a}(t)}{dt} = & \tilde{\ene}_{a}(t)\tilde{a}(t) + \sum_{k}V_{ak}(t) \exp \lsb -i\ek(t-t_{0}) \rsb \ca(t_{0}) \\
		- & i \sum_{k}\int_{t_{0}}^{t} dt' \exp\lsb -i\ek(t-t') \rsb V_{ak}(t)V_{ak}^{*}(t')\tilde{a}(t').
	\end{aligned}
	\label{eq:aheom1}
\end{equation}
The last term in \eqref{eq:aheom1} can be evaluated with the aid of the wide band limit resulting to
\begin{equation}
	-i \sum_{k}\int_{t_{0}}^{t} dt' \exp\lsb -i\ek(t-t') \rsb V_{ak}(t)V_{ak}^{*}(t')\tilde{a}(t') = -i\Delta(t)\tilde{a}(t),
	\label{eq:aheom1last}
\end{equation}
which then yields
\begin{equation}
	i\frac{d \tilde{a}(t)}{dt} = \lsb \tilde{\ene}_{a}(t) -i\Delta(t) \rsb \tilde{a}(t) + \sum_{k}V_{ak}(t) \exp \lsb -i\ek(t-t_{0}) \rsb \ca(t_{0}) 
	\label{eq:aheom2}
\end{equation}
This can be integrated to give
\begin{equation}
	\begin{aligned}
		\tilde{a}(t) = & \exp \lcb -i \int_{t_{0}}^{t} dt' \lsb \tilde{\ene}_{a}(t') -i\Delta(t') \rsb \rcb \tilde{a}(t_{0}) \
	+ \int_{t_{0}}^{t} dt' \exp \lcb -i \int_{t'}^{t} d\tau \lsb \tilde{\ene}_{a}(\tau) -i\Delta(\tau) \rsb \rcb \\
	\times & \sum_{k}V_{ak}(t') \exp \lsb -i\ek(t'-t_{0}) \rsb \ca(t_{0}). 
\end{aligned}
	\label{eq:at}
\end{equation}
Inserting $1=e^{-\lambda_{q}(b{\dagger}_{q}-b_{q})}e^{\lambda_{q}(b^{\dagger}_{q}-b_{q})}\
\equiv X(t)X^{\dagger}(t)$  
into the second term of \eqref{eq:at} yields
\begin{equation}
	\begin{aligned}
		\tilde{a}(t) = & \exp \lcb -i \int_{t_{0}}^{t} dt' \lsb \tilde{\ene}_{a}(t') -i\Delta(t') \rsb \rcb \tilde{a}(t_{0}) \
	+ \int_{t_{0}}^{t} dt' \exp \lcb -i \int_{t'}^{t} d\tau \lsb \tilde{\ene}_{a}(\tau) -i\Delta(\tau) \rsb \rcb \\
	\times & \sum_{k}\tilde{V}_{ak}(t')X(t') \exp \lsb -i\ek(t'-t_{0}) \rsb \ca(t_{0}). 
\end{aligned}
	\label{eq:at_insert}
\end{equation}
As in the Keldysh formalism, we 
take $\tilde{V}_{ak}(t')\approx V_{ak}(t') \langle X^{\dagger}(t') \rangle \equiv \bar{V}_{ak}(t')$
for $V_{ak} \ll \lambda$. At this point, we emphasize the difference between $\tilde{V}$ and $\bar{V}$.
With this, $\tilde{a}(t)$ takes the form
\begin{equation}
	\begin{aligned}
		\tilde{a}(t) = & \exp \lcb -i \int_{t_{0}}^{t} dt' \lsb \tilde{\ene}_{a}(t') -i\Delta(t') \rsb \rcb \tilde{a}(t_{0}) \
	+  \int_{t_{0}}^{t} dt' \exp \lcb -i \int_{t'}^{t} d\tau \lsb \tilde{\ene}_{a}(\tau) -i\Delta(\tau) \rsb \rcb \\
	\times & \sum_{k}\bar{V}_{ak}(t')X(t') \exp \lsb -i\ek(t'-t_{0}) \rsb \ca(t_{0}). 
\end{aligned}
	\label{eq:at_insert1}
\end{equation}
Multiplying both sides of \eqref{eq:at_insert1} with $X^{\dagger}(t)$ and 
using the decoupling approximation in \eqref{eq:gretad}, the expectation values 
with respect to the bath phonons gives
\begin{equation}
	\begin{aligned}
		a(t)= & \exp \lcb -i \int_{t_{0}}^{t} dt' \lsb \tilde{\ene}_{a}(t') -i\Delta(t') \rsb \rcb a(t_{0}) \
	\langle X(t_{0})X^{\dagger}(t) \rangle   \\
	+ & \int_{t_{0}}^{t} dt' \exp \lcb -i \int_{t'}^{t} d\tau \lsb \tilde{\ene}_{a}(\tau) -i\Delta(\tau) \rsb \rcb \\
	\times & \sum_{k}\bar{V}_{ak}(t') B(t'-t) \exp \lsb -i\ek(t'-t_{0}) \rsb \ca(t_{0}). 
\end{aligned}
	\label{eq:at_insert2}
\end{equation}
Consequently, the adsorbate electron's creation operator is obtained as
\begin{equation}
	\begin{aligned}
		\ac(t)= & \exp \lcb i \int_{t_{0}}^{t} dt' \lsb \tilde{\ene}_{a}(t') +i\Delta(t') \rsb \rcb \ac(t_{0}) \
	\langle X^{\dagger}(t_{0})X(t) \rangle   \\
	+ & \int_{t_{0}}^{t} dt' \exp \lcb i \int_{t'}^{t} d\tau \lsb \tilde{\ene}_{a}(\tau) +i\Delta(\tau) \rsb \rcb \\
	\times & \sum_{k}\bar{V}_{ak}^{*}(t') B(t-t')\exp \lsb i\ek(t'-t_{0}) \rsb \cc(t_{0}).
\end{aligned}
	\label{eq:act}
\end{equation}
Using \eqref{eq:at_insert2} and \eqref{eq:act}, the occupation number of the adsorbate is obtained 
from the expectation values of the adsorbate electron operators
\begin{equation}
	\langle n_{a}(t) \rangle = \langle \ac(t)a(t) \rangle.
	\label{eq:naeom}
\end{equation}
In evaluating \eqref{eq:naeom}, the terms proportional to $\langle \ac(t_{0})\ca(t_{0}) \rangle$ and
$\langle \cc(t_{0})a(t_{0}) \rangle$ are equal to zero since the adsorbate and 
metal electrodes electrons are assumed to be decoupled at a distant past ($t_{0}\rightarrow -\infty$).
After some algebraic manipulations and 
changing the summation over $k$ into an integration over $\ene$, we recover \eqref{eq:occu} as,
\begin{equation}
		\langle n_{a}(t) \rangle = \int d\ene f(\ene) \left|\int_{t_{0}}^{t}dt' \sqrt{\frac{\Delta(t')}{\pi}} \
		 B(t')
		 \exp \lcb -i\int_{t'}^{t}d\tau\lsb E_{a}(\tau) -\ene \rsb \rcb \right|^{2}.
	\label{eq:naeom2}
\end{equation}
where $ f(\ek)=\langle \cc(t_{0})\ca(t_{0}) \rangle$ is Fermi-Dirac distribution of the metal electrons,
and $E_{a}(t)\equiv\tilde{\ene}_{a}(t) -i\Delta(t)$ . 
To this end, we neglected the product of the first terms of \eqref{eq:act} and \eqref{eq:at_insert2} 
which is a rapidly decaying transient.  

\section{Numerical Evaluation}
\label{appendix:numerical}
In the position representation, the occupation number can be written as
\begin{equation}
	\langle n_{a}(z) \rangle =  \frac{1}{v^{2}}\int d\ene f(\ene) \left|p(\ene,z)\right|^2.
	\label{eq:occu1}
\end{equation}
$p(\ene,z)$ is defined as
\begin{equation}
	p(\ene,z) =  \int_{z_{0}}^{z} dz'\sqrt{\frac{\Delta(z')}{\pi}} \
	B(z')\exp\left\{-\frac{i}{v}\int_{z'}^{z} dz'' \left[\tilde{\ene}_{a}(z'') -i\Delta(z'') -\ene \right]\right\},
	\label{eq:pez}
\end{equation}
where we replaced the time dependence in the bath correlation function with $z$.
Since $B(t)$ is maximum when $t=0$ and approaches $1$ when $t\rightarrow \infty$,
we shift $B(z)$ by $z+|z_{min}|$ with $z_{min} = -20$ Bohr being the starting point in the 
incoming portion of the trajectory to account for this behavior.
We do not consider adsorption and assume that the parameters 
$\Delta(z)$ and $\tilde{\ene}_{a}(z)$ can be expressed as Gaussian functions (see \eqref{eq:params}).

\begin{figure}
\centering
\captionsetup{justification=raggedright,singlelinecheck=false}
\subfloat[\centering]{
	\includegraphics[scale=0.5]{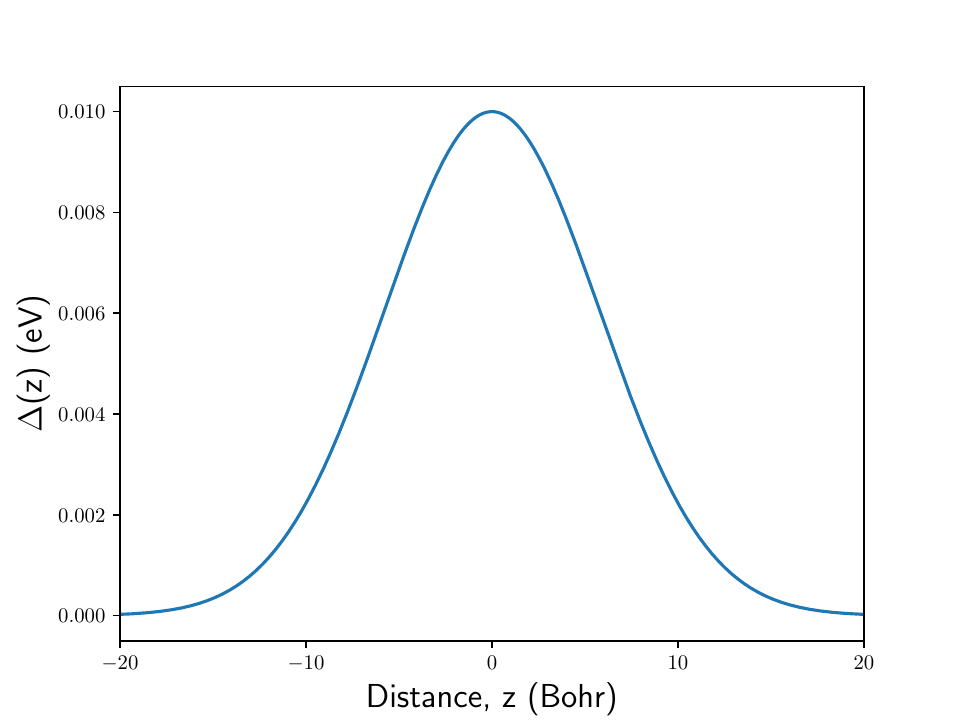} \label{fig:delta}
}
\qquad
\subfloat[\centering]{
	\includegraphics[scale=0.5]{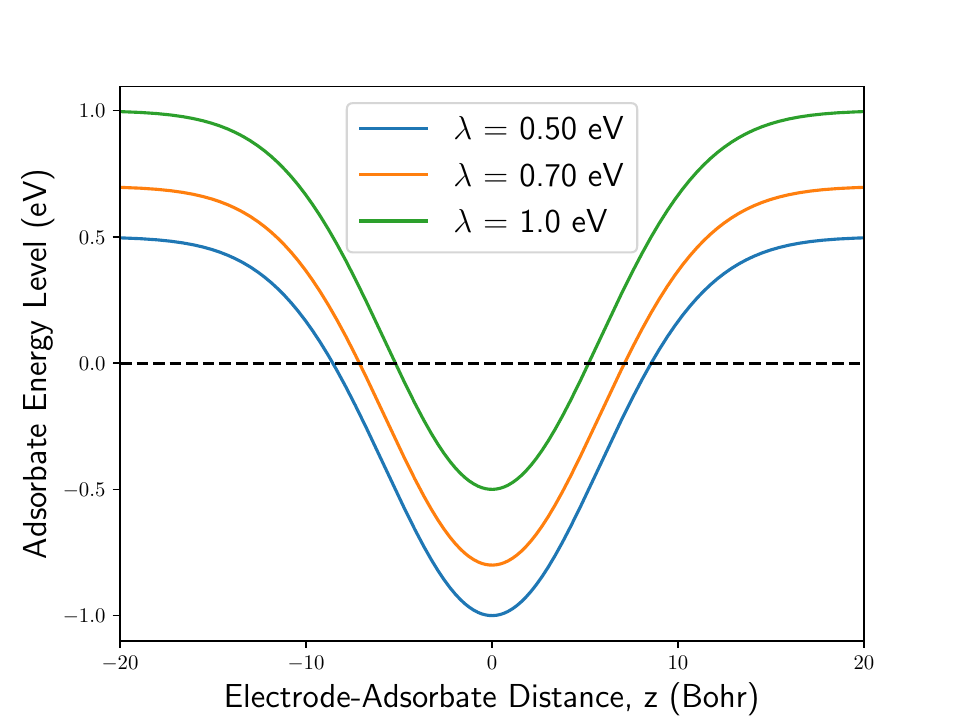} \label{fig:en_level}
}
\caption{\label{fig:parameters} (a) The adorbate level resonance width $\Delta(z)$ for $\Delta=0.01~$eV and 
	$\alpha = 0.015~\alpha_{0}^{-2}$. (b) The adsorbate energy level $\tilde{\ene}_{a}(z)$ for different
	strengths of reorganization energy $\lambda$.
	Other parameters are $\ene_{F}=0.0~$eV, $\kappa = \alpha$, $\ene_{0} = -1.5~$eV and $\ene_{\infty}=\ene_{F}$.
	The dashed black line corresponds to the electrode's Fermi level.
}
\end{figure}
\figref{fig:parameters} shows the resonance width and the energy level of adsorbate as functions
of $z$. Our chosen parameters correspond to a system consisting of an initially empty adsorbate orbital 
with an energy $\ene_{\infty}$ that is shifted upwards by the reorganization energy due to e-ph coupling. As the adsorbate
approaches near the electrode, the energy level starts to be filled when $\tilde{\ene}_{a}(z)=\ene_{F}$ and broadened by
a magnitude $\Delta$ until $z=0.0$ Bohr where it is at maximum. 
As the adsorbate moves away from the electrode, this level gradually becomes unnocuppied again and returns
to its value $\ene_{\infty}$ as $z\rightarrow \infty$.

We numerically evaluate \eqref{eq:occu1} by
taking the derivative of \eqref{eq:pez} and integrating the resulting first-order complex differential
equation, 
\begin{equation}
	\frac{dp(\ene,z)}{dz}=\sqrt{\frac{\Delta(z)}{\pi}}B(z)-\frac{i}{v} \left[\tilde{\ene}_{a}(z)-i\Delta(z) -\ene \right]p(\ene,z).
	\label{eq:de}
\end{equation}
We assume that the electron transfer occurs at $T=300$ K and
integrate \eqref{eq:de} over the trajectory that starts at $z=-20~$Bohr and ends at $z=20~$Bohr 
with the interval $\Delta z = 0.01~$Bohr using the complex ODE module of PYTHON. The initial condition
corresponds to an empty adsorbate orbital at $z=-20$ Bohr.
The energy integration is performed from $-10$ eV to $10$ eV with a grid size of $0.001$ eV.
We consider different adsorbate velocities which includes $v=0.23~\alpha_{0}$eV, $v=0.1~\alpha_{0}$eV, $v=0.01~\alpha_{0}$eV,
and $v=0.001~\alpha_{0}$eV. $v=0.23~\alpha_{0}$eV is the thermal velocity of an adsorbate with mass $m=1$
a kinetic energy equal to the thermal energy. In the absence of acceleration induced internally 
by the potential energy landscape of the adsorbate path and other external forces, this velocity 
may represent an upper limit in our study. The corresponding round trip simulation times $t=\frac{\hbar |z|}{v}$
for our chosen velocities are $114.5$ fs, $263.3$ fs, $2.633\times10^{3}$ fs, and $2.633\times10^{4}$ fs, respectively. 
As one may expect, the correlation function $B(t)$ requires a significant number of Matsubara
frequencies $N_{k}\approx 10^{3}$ at low temperatures and can significantly slow down the calculations. 
However, when the electron transfer occurs at room temperature as is the case here, we found that 
it is sufficient to use $N_{k}=2$.

\section{Marcus Theory Limit}
\label{appendix:marcus}
The electron transfer rate in the Marcus theory limit can be derived by first 
performing Fourier transform of the Lorentzian PDOS in \eqref{eq:nad} and
writing the rate as
\begin{equation}
	k = \frac{2\Delta}{\pi}\int d\ene f(\ene) \int_{-\infty}^{\infty}dt e^{-\Delta|t|}e^{-i\left(\tilde{\ene}_{a}-\ene \right)t} .
	\label{eq:marcusrate}
\end{equation}
Performing the thermal average as discussed in\cite{Mohr2000,Schmickler1993} of \eqref{eq:marcusrate} with respect to bath phonons results in
\begin{equation}
	k = \frac{2\Delta}{\pi}\int d\ene f(\ene) \int_{-\infty}^{\infty}dt e^{-\Delta|t|}e^{-i\left(\ene_{a}+\lambda-\ene \right)t}e^{-\frac{\lambda}{\beta}t^{2}} .
	\label{eq:marcusrate2}
\end{equation}
Marcus theory assumes that $k_{B}T\gg \Delta$ which is equivalent to neglecting 
$\Delta$ in the exponential in \eqref{eq:marcusrate2}.
Evaluating the time integral leads to the familiar Gaussian form of $k$ as
\begin{equation}
	k = \frac{2\Delta}{\pi}\int d\ene f(\ene) \sqrt{\frac{\pi}{4\lambda k_{B}T}}\exp\lcb-\frac{\lsb \ene_{a}+\lambda-\ene\rsb^{2}}{4\lambda k_{B}T}\rcb.
	\label{eq:marcusrate3}
\end{equation}

\section{Derivation of Energy Exchange Rate}
\label{appendix:en_ex_rate}
In order to arrive at \eqref{eq:en_tr}, we first take the time derivative of
\eqref{eq:lf_totham} giving
\begin{equation}
	\dot{\tilde{H}} = \dot{\tilde{\ene}}_{a}(t)\ac(t)a(t)+\sum_{k}\
	\lsb \dot{V}_{ak}(t)X^{\dagger}(t)\ac(t)\ca(t)+h.c.\rsb.
	\label{eq:deriv_lf_totham}
\end{equation}
To this end, the corresponding equations of motion for each operators 
have been used. After some algebraic manipulations, \eqref{eq:deriv_lf_totham}
may be expressed as
\begin{equation}
	\dot{\tilde{H}} = \dot{\tilde{\ene}}_{a}(t)\ac(t)a(t)+\frac{\dot{\Delta}(t)}{2\Delta(t)}\sum_{k}\
	\lsb V_{ak}(t)X^{\dagger}(t)\ac(t)\ca(t)+h.c.\rsb.
	\label{eq:deriv_lf_totham1}
\end{equation}
From the equations of motion of $\ac$ and $\ca$,
\begin{equation}
	\ac(t)\ca(t) = \frac{1}{V_{ak}(t)X^{\dagger}(t)}\lsb i\ac(t)\dot{a}(t)-\tilde{\ene}_{a}(t)\ac(t)a(t) \rsb,
	\label{eq:adag_ck}
\end{equation}
which yields
\begin{equation}
	\dot{\tilde{H}} = \dot{\tilde{\ene}}_{a}(t)\ac(t)a(t)-\frac{\dot{\Delta}(t)}{\Delta(t)}\tilde{\ene}_{a}(t)\ac(t)a(t)\
	+\frac{\dot{\Delta}(t)}{\Delta(t)}\Re\lsb i\ac(t)\dot{a}(t)\rsb.
	\label{eq:deriv_lf_totham2}
\end{equation}
Using \eqref{eq:act} and the derivative of \eqref{eq:at_insert2},
\begin{equation}
	\begin{aligned}
		i\ac(t)\dot{a}(t) &= E_{a}(t)\ac(t)a(t) + i\sum_{k}\cc(t_{0})\ca(t_{0})|V_{ak}|^{2}\int_{t_{0}}^{t}dt'u(t)u(t')X(t)X^{\dagger}(t) \\
		&\times\exp\lcb i\lsb\int_{t'}^{t}E_{a}^{*}(t'')dt''+\ene_{k}(t'-t) \rsb\rcb +\
		\exp\lsb i\int_{t_{0}}^{t}E_{a}^{*}(t')dt' \rsb\ac(t_{0})\ca(t_{0}),
\end{aligned}
	\label{eq:adag_dota}
\end{equation}
Substituting \eqref{eq:adag_dota} into \eqref{eq:deriv_lf_totham2} and taking the expectation value,
we obtain an expression for the energy transfer rate as
\begin{equation}
	\begin{aligned}
	\dot{\mathcal{E}} &\equiv \langle \dot{\tilde{H}} \rangle = \dot{\tilde{\ene}}_{a}(t)\langle n_{a}(t) \rangle \
	+\frac{\dot{\Delta}(t)}{\Delta(t)}\Re\lsb i\int d\ene f(\ene) \sqrt{\frac{\Delta(t)}{\pi}}p^{*}(\ene,t)\rsb, \\
	& = \dot{\tilde{\ene}}_{a}(t)\langle n_{a}(t) \rangle \
	+\frac{\dot{\Delta}(t)}{\sqrt{\pi\Delta(t)}}\int d\ene f(\ene) \Im \lsb p(\ene,t)\rsb
\end{aligned}
	\label{eq:ap_en_tran}
\end{equation}
with
\begin{equation}
	p(\ene,t) \equiv \int_{t_{0}}^{t}dt'\sqrt{\frac{\Delta(t')}{\pi}}B(t')\exp\lcb i\int_{t'}^{t}dt''\lsb E_{a}(t'')-\ene \rsb\rcb.
	\label{eq:ap_p}
\end{equation}
To this end, we note that the expectation value of the last term of \eqref{eq:adag_dota} is zero for $t_{0}=-\infty$.

\bibliography{library.bib}

\begin{thebibliography}{49}%
\makeatletter
\providecommand \@ifxundefined [1]{%
 \@ifx{#1\undefined}
}%
\providecommand \@ifnum [1]{%
 \ifnum #1\expandafter \@firstoftwo
 \else \expandafter \@secondoftwo
 \fi
}%
\providecommand \@ifx [1]{%
 \ifx #1\expandafter \@firstoftwo
 \else \expandafter \@secondoftwo
 \fi
}%
\providecommand \natexlab [1]{#1}%
\providecommand \enquote  [1]{``#1''}%
\providecommand \bibnamefont  [1]{#1}%
\providecommand \bibfnamefont [1]{#1}%
\providecommand \citenamefont [1]{#1}%
\providecommand \href@noop [0]{\@secondoftwo}%
\providecommand \href [0]{\begingroup \@sanitize@url \@href}%
\providecommand \@href[1]{\@@startlink{#1}\@@href}%
\providecommand \@@href[1]{\endgroup#1\@@endlink}%
\providecommand \@sanitize@url [0]{\catcode `\\12\catcode `\$12\catcode
  `\&12\catcode `\#12\catcode `\^12\catcode `\_12\catcode `\%12\relax}%
\providecommand \@@startlink[1]{}%
\providecommand \@@endlink[0]{}%
\providecommand \url  [0]{\begingroup\@sanitize@url \@url }%
\providecommand \@url [1]{\endgroup\@href {#1}{\urlprefix }}%
\providecommand \urlprefix  [0]{URL }%
\providecommand \Eprint [0]{\href }%
\providecommand \doibase [0]{http://dx.doi.org/}%
\providecommand \selectlanguage [0]{\@gobble}%
\providecommand \bibinfo  [0]{\@secondoftwo}%
\providecommand \bibfield  [0]{\@secondoftwo}%
\providecommand \translation [1]{[#1]}%
\providecommand \BibitemOpen [0]{}%
\providecommand \bibitemStop [0]{}%
\providecommand \bibitemNoStop [0]{.\EOS\space}%
\providecommand \EOS [0]{\spacefactor3000\relax}%
\providecommand \BibitemShut  [1]{\csname bibitem#1\endcsname}%
\let\auto@bib@innerbib\@empty
\bibitem [{\citenamefont {Marcus}(1965)}]{Marcus1965}%
  \BibitemOpen
  \bibfield  {author} {\bibinfo {author} {\bibfnamefont {R.~A.}\ \bibnamefont
  {Marcus}},\ }\href@noop {} {\bibfield  {journal} {\bibinfo  {journal} {The
  Journal of Chemical Physics}\ }\textbf {\bibinfo {volume} {43}},\ \bibinfo
  {pages} {679} (\bibinfo {year} {1965})}\BibitemShut {NoStop}%
\bibitem [{\citenamefont {Hush}(1958)}]{Hush1958}%
  \BibitemOpen
  \bibfield  {author} {\bibinfo {author} {\bibfnamefont {N.~S.}\ \bibnamefont
  {Hush}},\ }\href@noop {} {\bibfield  {journal} {\bibinfo  {journal} {The
  Journal of Chemical Physics}\ }\textbf {\bibinfo {volume} {28}},\ \bibinfo
  {pages} {962} (\bibinfo {year} {1958})}\BibitemShut {NoStop}%
\bibitem [{\citenamefont {Levich}(1970)}]{Levich1970}%
  \BibitemOpen
  \bibfield  {author} {\bibinfo {author} {\bibfnamefont {V.}~\bibnamefont
  {Levich}},\ }\href@noop {} {\emph {\bibinfo {title} {Physical Chemistry. An
  Advanced Treatise}}},\ edited by\ \bibinfo {editor} {\bibfnamefont
  {H.}~\bibnamefont {Eyring}}, \bibinfo {editor} {\bibfnamefont
  {D.}~\bibnamefont {Henderson}}, \ and\ \bibinfo {editor} {\bibfnamefont
  {W.}~\bibnamefont {Jost}},\ Vol.~\bibinfo {volume} {Xb}\ (\bibinfo
  {publisher} {Academic Press},\ \bibinfo {year} {1970})\BibitemShut {NoStop}%
\bibitem [{\citenamefont {Schmickler}(1986)}]{Schmickler1986}%
  \BibitemOpen
  \bibfield  {author} {\bibinfo {author} {\bibfnamefont {W.}~\bibnamefont
  {Schmickler}},\ }\href@noop {} {\bibfield  {journal} {\bibinfo  {journal} {J.
  Electroanal. Chem}\ }\textbf {\bibinfo {volume} {204}},\ \bibinfo {pages}
  {31} (\bibinfo {year} {1986})}\BibitemShut {NoStop}%
\bibitem [{\citenamefont {Hewson}\ and\ \citenamefont
  {Newns}(1974)}]{Hewson1974}%
  \BibitemOpen
  \bibfield  {author} {\bibinfo {author} {\bibfnamefont {A.~C.}\ \bibnamefont
  {Hewson}}\ and\ \bibinfo {author} {\bibfnamefont {D.~M.}\ \bibnamefont
  {Newns}},\ }\href {https://iopscience.iop.org/article/10.7567/JJAPS.2S2.121}
  {\bibfield  {journal} {\bibinfo  {journal} {Japanese Journal of Applied
  Physics}\ }\textbf {\bibinfo {volume} {13}},\ \bibinfo {pages} {121}
  (\bibinfo {year} {1974})}\BibitemShut {NoStop}%
\bibitem [{\citenamefont {Citrin}\ and\ \citenamefont
  {Hamann}(1977)}]{Citrin1977}%
  \BibitemOpen
  \bibfield  {author} {\bibinfo {author} {\bibfnamefont {P.~H.}\ \bibnamefont
  {Citrin}}\ and\ \bibinfo {author} {\bibfnamefont {D.~R.}\ \bibnamefont
  {Hamann}},\ }\href@noop {} {\bibfield  {journal} {\bibinfo  {journal}
  {Physical Review B}\ }\textbf {\bibinfo {volume} {15}},\ \bibinfo {pages}
  {2923} (\bibinfo {year} {1977})}\BibitemShut {NoStop}%
\bibitem [{\citenamefont {Song}\ and\ \citenamefont {Marcus}(1993)}]{Song1993}%
  \BibitemOpen
  \bibfield  {author} {\bibinfo {author} {\bibfnamefont {X.}~\bibnamefont
  {Song}}\ and\ \bibinfo {author} {\bibfnamefont {R.~A.}\ \bibnamefont
  {Marcus}},\ }\href@noop {} {\bibfield  {journal} {\bibinfo  {journal} {The
  Journal of Chemical Physics}\ }\textbf {\bibinfo {volume} {99}},\ \bibinfo
  {pages} {7768} (\bibinfo {year} {1993})}\BibitemShut {NoStop}%
\bibitem [{\citenamefont {Sebastian}(1989)}]{Sebastian1989}%
  \BibitemOpen
  \bibfield  {author} {\bibinfo {author} {\bibfnamefont {K.~L.}\ \bibnamefont
  {Sebastian}},\ }\href@noop {} {\bibfield  {journal} {\bibinfo  {journal} {The
  Journal of Chemical Physics}\ }\textbf {\bibinfo {volume} {90}},\ \bibinfo
  {pages} {5056} (\bibinfo {year} {1989})}\BibitemShut {NoStop}%
\bibitem [{\citenamefont {Smith}\ and\ \citenamefont
  {Hynes}(1993)}]{Smith1993}%
  \BibitemOpen
  \bibfield  {author} {\bibinfo {author} {\bibfnamefont {B.~B.}\ \bibnamefont
  {Smith}}\ and\ \bibinfo {author} {\bibfnamefont {J.~T.}\ \bibnamefont
  {Hynes}},\ }\href@noop {} {\bibfield  {journal} {\bibinfo  {journal} {The
  Journal of Chemical Physics}\ }\textbf {\bibinfo {volume} {99}},\ \bibinfo
  {pages} {6517} (\bibinfo {year} {1993})}\BibitemShut {NoStop}%
\bibitem [{\citenamefont {Mohr}\ and\ \citenamefont
  {Schmickler}(2000)}]{Mohr2000}%
  \BibitemOpen
  \bibfield  {author} {\bibinfo {author} {\bibfnamefont {J.-H.}\ \bibnamefont
  {Mohr}}\ and\ \bibinfo {author} {\bibfnamefont {W.}~\bibnamefont
  {Schmickler}},\ }\href@noop {} {\bibfield  {journal} {\bibinfo  {journal}
  {Physical Review Letters}\ }\textbf {\bibinfo {volume} {84}},\ \bibinfo
  {pages} {1051} (\bibinfo {year} {2000})}\BibitemShut {NoStop}%
\bibitem [{\citenamefont {Tanaka}\ and\ \citenamefont
  {Hsu}(1999)}]{Tanaka1999}%
  \BibitemOpen
  \bibfield  {author} {\bibinfo {author} {\bibfnamefont {S.}~\bibnamefont
  {Tanaka}}\ and\ \bibinfo {author} {\bibfnamefont {C.~P.}\ \bibnamefont
  {Hsu}},\ }\href@noop {} {\bibfield  {journal} {\bibinfo  {journal} {Journal
  of Chemical Physics}\ }\textbf {\bibinfo {volume} {111}},\ \bibinfo {pages}
  {11117} (\bibinfo {year} {1999})}\BibitemShut {NoStop}%
\bibitem [{\citenamefont {Wodtke}\ \emph {et~al.}(2004)\citenamefont {Wodtke},
  \citenamefont {Tully},\ and\ \citenamefont {Auerbach}}]{Wodtke2004}%
  \BibitemOpen
  \bibfield  {author} {\bibinfo {author} {\bibfnamefont {A.~M.}\ \bibnamefont
  {Wodtke}}, \bibinfo {author} {\bibfnamefont {J.~C.}\ \bibnamefont {Tully}}, \
  and\ \bibinfo {author} {\bibfnamefont {D.~J.}\ \bibnamefont {Auerbach}},\
  }\href@noop {} {\bibfield  {journal} {\bibinfo  {journal} {International
  Reviews in Physical Chemistry}\ }\textbf {\bibinfo {volume} {23}},\ \bibinfo
  {pages} {513} (\bibinfo {year} {2004})}\BibitemShut {NoStop}%
\bibitem [{\citenamefont {Piechota}\ and\ \citenamefont
  {Meyer}(2019)}]{Piechota2019}%
  \BibitemOpen
  \bibfield  {author} {\bibinfo {author} {\bibfnamefont {E.~J.}\ \bibnamefont
  {Piechota}}\ and\ \bibinfo {author} {\bibfnamefont {G.~J.}\ \bibnamefont
  {Meyer}},\ }\href@noop {} {\bibfield  {journal} {\bibinfo  {journal} {Journal
  of Chemical Education}\ }\textbf {\bibinfo {volume} {96}},\ \bibinfo {pages}
  {2450} (\bibinfo {year} {2019})}\BibitemShut {NoStop}%
\bibitem [{\citenamefont {Brako}\ and\ \citenamefont
  {Newns}(1981{\natexlab{a}})}]{Brako1981}%
  \BibitemOpen
  \bibfield  {author} {\bibinfo {author} {\bibfnamefont {R.}~\bibnamefont
  {Brako}}\ and\ \bibinfo {author} {\bibfnamefont {D.~M.}\ \bibnamefont
  {Newns}},\ }\href@noop {} {\bibfield  {journal} {\bibinfo  {journal} {Surface
  Science}\ }\textbf {\bibinfo {volume} {108}},\ \bibinfo {pages} {253}
  (\bibinfo {year} {1981}{\natexlab{a}})}\BibitemShut {NoStop}%
\bibitem [{\citenamefont {Blandin}\ \emph {et~al.}(1976)\citenamefont
  {Blandin}, \citenamefont {Nourtier},\ and\ \citenamefont
  {Hone}}]{Blandin1976}%
  \BibitemOpen
  \bibfield  {author} {\bibinfo {author} {\bibfnamefont {A.}~\bibnamefont
  {Blandin}}, \bibinfo {author} {\bibfnamefont {A.}~\bibnamefont {Nourtier}}, \
  and\ \bibinfo {author} {\bibfnamefont {D.}~\bibnamefont {Hone}},\ }\href@noop
  {} {\bibfield  {journal} {\bibinfo  {journal} {Journal de Physique}\ }\textbf
  {\bibinfo {volume} {37}},\ \bibinfo {pages} {369} (\bibinfo {year}
  {1976})}\BibitemShut {NoStop}%
\bibitem [{\citenamefont {Yoshimori}\ \emph {et~al.}(1984)\citenamefont
  {Yoshimori}, \citenamefont {Kawai},\ and\ \citenamefont
  {Makoshi}}]{Yoshimori1984}%
  \BibitemOpen
  \bibfield  {author} {\bibinfo {author} {\bibfnamefont {A.}~\bibnamefont
  {Yoshimori}}, \bibinfo {author} {\bibfnamefont {H.}~\bibnamefont {Kawai}}, \
  and\ \bibinfo {author} {\bibfnamefont {K.}~\bibnamefont {Makoshi}},\
  }\href@noop {} {\bibfield  {journal} {\bibinfo  {journal} {Progress of
  Theoretical Physics Supplement}\ }\textbf {\bibinfo {volume} {80}},\ \bibinfo
  {pages} {203} (\bibinfo {year} {1984})}\BibitemShut {NoStop}%
\bibitem [{\citenamefont {Kasai}\ and\ \citenamefont
  {Okiji}(1987)}]{Kasai1987}%
  \BibitemOpen
  \bibfield  {author} {\bibinfo {author} {\bibfnamefont {H.}~\bibnamefont
  {Kasai}}\ and\ \bibinfo {author} {\bibfnamefont {A.}~\bibnamefont {Okiji}},\
  }\href@noop {} {\bibfield  {journal} {\bibinfo  {journal} {Surface Science}\
  }\textbf {\bibinfo {volume} {183}},\ \bibinfo {pages} {147} (\bibinfo {year}
  {1987})}\BibitemShut {NoStop}%
\bibitem [{\citenamefont {Yoshimori}\ and\ \citenamefont
  {Makoshi}(1986)}]{Yoshimori1986}%
  \BibitemOpen
  \bibfield  {author} {\bibinfo {author} {\bibfnamefont {A.}~\bibnamefont
  {Yoshimori}}\ and\ \bibinfo {author} {\bibfnamefont {K.}~\bibnamefont
  {Makoshi}},\ }\href@noop {} {\bibfield  {journal} {\bibinfo  {journal}
  {Progress in Surface Science}\ }\textbf {\bibinfo {volume} {21}},\ \bibinfo
  {pages} {251} (\bibinfo {year} {1986})}\BibitemShut {NoStop}%
\bibitem [{\citenamefont {Mizielinski}\ \emph {et~al.}(2005)\citenamefont
  {Mizielinski}, \citenamefont {Bird}, \citenamefont {Persson},\ and\
  \citenamefont {Holloway}}]{Mizielinski2005}%
  \BibitemOpen
  \bibfield  {author} {\bibinfo {author} {\bibfnamefont {M.~S.}\ \bibnamefont
  {Mizielinski}}, \bibinfo {author} {\bibfnamefont {D.~M.}\ \bibnamefont
  {Bird}}, \bibinfo {author} {\bibfnamefont {M.}~\bibnamefont {Persson}}, \
  and\ \bibinfo {author} {\bibfnamefont {S.}~\bibnamefont {Holloway}},\
  }\href@noop {} {\bibfield  {journal} {\bibinfo  {journal} {Journal of
  Chemical Physics}\ }\textbf {\bibinfo {volume} {122}} (\bibinfo {year}
  {2005})}\BibitemShut {NoStop}%
\bibitem [{\citenamefont {Brako}\ and\ \citenamefont
  {Newns}(1980)}]{Brako1980}%
  \BibitemOpen
  \bibfield  {author} {\bibinfo {author} {\bibfnamefont {R.}~\bibnamefont
  {Brako}}\ and\ \bibinfo {author} {\bibfnamefont {D.~M.}\ \bibnamefont
  {Newns}},\ }\href@noop {} {\bibfield  {journal} {\bibinfo  {journal} {Solid
  State Communications}\ }\textbf {\bibinfo {volume} {33}},\ \bibinfo {pages}
  {713} (\bibinfo {year} {1980})}\BibitemShut {NoStop}%
\bibitem [{\citenamefont {Plihal}\ and\ \citenamefont
  {Langreth}(2018)}]{Plihal1998}%
  \BibitemOpen
  \bibfield  {author} {\bibinfo {author} {\bibfnamefont {M.}~\bibnamefont
  {Plihal}}\ and\ \bibinfo {author} {\bibfnamefont {D.~C.}\ \bibnamefont
  {Langreth}},\ }\href@noop {} {\bibfield  {journal} {\bibinfo  {journal}
  {Journal of Chemical Physics}\ }\textbf {\bibinfo {volume} {149}} (\bibinfo
  {year} {2018})}\BibitemShut {NoStop}%
\bibitem [{\citenamefont {Newns}(1986)}]{Newns1986}%
  \BibitemOpen
  \bibfield  {author} {\bibinfo {author} {\bibfnamefont {D.~M.}\ \bibnamefont
  {Newns}},\ }\href@noop {} {\bibfield  {journal} {\bibinfo  {journal} {Surface
  Science}\ }\textbf {\bibinfo {volume} {171}},\ \bibinfo {pages} {600}
  (\bibinfo {year} {1986})}\BibitemShut {NoStop}%
\bibitem [{\citenamefont {Kasai}\ and\ \citenamefont
  {Okiji}(1991)}]{Kasai1991}%
  \BibitemOpen
  \bibfield  {author} {\bibinfo {author} {\bibfnamefont {H.}~\bibnamefont
  {Kasai}}\ and\ \bibinfo {author} {\bibfnamefont {A.}~\bibnamefont {Okiji}},\
  }\href@noop {} {\bibfield  {journal} {\bibinfo  {journal} {Surface Science}\
  }\textbf {\bibinfo {volume} {242}},\ \bibinfo {pages} {394} (\bibinfo {year}
  {1991})}\BibitemShut {NoStop}%
\bibitem [{\citenamefont {Gross}\ and\ \citenamefont
  {Brenig}(1993)}]{Gross1993}%
  \BibitemOpen
  \bibfield  {author} {\bibinfo {author} {\bibfnamefont {A.}~\bibnamefont
  {Gross}}\ and\ \bibinfo {author} {\bibfnamefont {W.}~\bibnamefont {Brenig}},\
  }\href@noop {} {\bibfield  {journal} {\bibinfo  {journal} {Chemical Physics}\
  }\textbf {\bibinfo {volume} {177}},\ \bibinfo {pages} {497} (\bibinfo {year}
  {1993})}\BibitemShut {NoStop}%
\bibitem [{\citenamefont {Lam}\ \emph {et~al.}(2019)\citenamefont {Lam},
  \citenamefont {Soudackov},\ and\ \citenamefont {Hammes-Schiffer}}]{Lam2019}%
  \BibitemOpen
  \bibfield  {author} {\bibinfo {author} {\bibfnamefont {Y.~C.}\ \bibnamefont
  {Lam}}, \bibinfo {author} {\bibfnamefont {A.~V.}\ \bibnamefont {Soudackov}},
  \ and\ \bibinfo {author} {\bibfnamefont {S.}~\bibnamefont
  {Hammes-Schiffer}},\ }\href@noop {} {\bibfield  {journal} {\bibinfo
  {journal} {Journal of Physical Chemistry Letters}\ }\textbf {\bibinfo
  {volume} {10}},\ \bibinfo {pages} {5312} (\bibinfo {year}
  {2019})}\BibitemShut {NoStop}%
\bibitem [{\citenamefont {Dou}\ and\ \citenamefont {Subotnik}(2020)}]{Dou2020}%
  \BibitemOpen
  \bibfield  {author} {\bibinfo {author} {\bibfnamefont {W.}~\bibnamefont
  {Dou}}\ and\ \bibinfo {author} {\bibfnamefont {J.}~\bibnamefont {Subotnik}},\
  }\href@noop {} {\bibfield  {journal} {\bibinfo  {journal} {Journal of
  Physical Chemistry A}\ }\textbf {\bibinfo {volume} {124}},\ \bibinfo {pages}
  {757} (\bibinfo {year} {2020})}\BibitemShut {NoStop}%
\bibitem [{\citenamefont {Hewson}\ and\ \citenamefont
  {Newns}(1980)}]{Hewson1980}%
  \BibitemOpen
  \bibfield  {author} {\bibinfo {author} {\bibfnamefont {A.~C.}\ \bibnamefont
  {Hewson}}\ and\ \bibinfo {author} {\bibfnamefont {D.~M.}\ \bibnamefont
  {Newns}},\ }\href@noop {} {\bibfield  {journal} {\bibinfo  {journal} {Journal
  of Physics C: Solid State Physics}\ }\textbf {\bibinfo {volume} {13}},\
  \bibinfo {pages} {4477} (\bibinfo {year} {1980})}\BibitemShut {NoStop}%
\bibitem [{\citenamefont {Chen}\ \emph {et~al.}(2005)\citenamefont {Chen},
  \citenamefont {Lü},\ and\ \citenamefont {Zhu}}]{Chen2005}%
  \BibitemOpen
  \bibfield  {author} {\bibinfo {author} {\bibfnamefont {Z.~Z.}\ \bibnamefont
  {Chen}}, \bibinfo {author} {\bibfnamefont {R.}~\bibnamefont {Lü}}, \ and\
  \bibinfo {author} {\bibfnamefont {B.~F.}\ \bibnamefont {Zhu}},\ }\href@noop
  {} {\bibfield  {journal} {\bibinfo  {journal} {Physical Review B - Condensed
  Matter and Materials Physics}\ }\textbf {\bibinfo {volume} {71}} (\bibinfo
  {year} {2005})}\BibitemShut {NoStop}%
\bibitem [{\citenamefont {Jauho}\ \emph {et~al.}(1994)\citenamefont {Jauho},
  \citenamefont {Wingreen},\ and\ \citenamefont {Meir}}]{Jauho1994}%
  \BibitemOpen
  \bibfield  {author} {\bibinfo {author} {\bibfnamefont {A.-P.}\ \bibnamefont
  {Jauho}}, \bibinfo {author} {\bibfnamefont {N.~S.}\ \bibnamefont {Wingreen}},
  \ and\ \bibinfo {author} {\bibfnamefont {Y.}~\bibnamefont {Meir}},\
  }\href@noop {} {\bibfield  {journal} {\bibinfo  {journal} {Physical Review
  B}\ }\textbf {\bibinfo {volume} {50}},\ \bibinfo {pages} {5528} (\bibinfo
  {year} {1994})}\BibitemShut {NoStop}%
\bibitem [{\citenamefont {Langreth}(1976)}]{Langreth1976}%
  \BibitemOpen
  \bibfield  {author} {\bibinfo {author} {\bibfnamefont {D.}~\bibnamefont
  {Langreth}},\ }\href@noop {} {\emph {\bibinfo {title} {Linear and Non Linear
  Electron Transport in Solids}}},\ \bibinfo {edition} {nato asi series b}\
  ed.,\ edited by\ \bibinfo {editor} {\bibfnamefont {J.~T.}\ \bibnamefont
  {Devreese}}\ and\ \bibinfo {editor} {\bibfnamefont {E.}~\bibnamefont {van
  Doren}},\ Vol.~\bibinfo {volume} {17}\ (\bibinfo  {publisher} {Plenum},\
  \bibinfo {year} {1976})\ p.~\bibinfo {pages} {3}\BibitemShut {NoStop}%
\bibitem [{\citenamefont {Odashima}\ and\ \citenamefont
  {Lewenkopf}(2017)}]{Odashima2017}%
  \BibitemOpen
  \bibfield  {author} {\bibinfo {author} {\bibfnamefont {M.~M.}\ \bibnamefont
  {Odashima}}\ and\ \bibinfo {author} {\bibfnamefont {C.~H.}\ \bibnamefont
  {Lewenkopf}},\ }\href@noop {} {\bibfield  {journal} {\bibinfo  {journal}
  {Physical Review B}\ }\textbf {\bibinfo {volume} {95}} (\bibinfo {year}
  {2017})}\BibitemShut {NoStop}%
\bibitem [{\citenamefont {Zhu}\ and\ \citenamefont {Balatsky}(2003)}]{Zhu2003}%
  \BibitemOpen
  \bibfield  {author} {\bibinfo {author} {\bibfnamefont {J.~X.}\ \bibnamefont
  {Zhu}}\ and\ \bibinfo {author} {\bibfnamefont {A.~V.}\ \bibnamefont
  {Balatsky}},\ }\href@noop {} {\bibfield  {journal} {\bibinfo  {journal}
  {Physical Review B}\ }\textbf {\bibinfo {volume} {67}} (\bibinfo {year}
  {2003})}\BibitemShut {NoStop}%
\bibitem [{\citenamefont {Tanimura}\ and\ \citenamefont
  {Kubo}(1989)}]{Tanimura1989}%
  \BibitemOpen
  \bibfield  {author} {\bibinfo {author} {\bibfnamefont {Y.}~\bibnamefont
  {Tanimura}}\ and\ \bibinfo {author} {\bibfnamefont {R.}~\bibnamefont
  {Kubo}},\ }\href@noop {} {\bibfield  {journal} {\bibinfo  {journal} {Journal
  of the Physical Society of Japan}\ }\textbf {\bibinfo {volume} {58}},\
  \bibinfo {pages} {101} (\bibinfo {year} {1989})}\BibitemShut {NoStop}%
\bibitem [{\citenamefont {Tanimura}(1990)}]{Tanimura1990}%
  \BibitemOpen
  \bibfield  {author} {\bibinfo {author} {\bibfnamefont {Y.}~\bibnamefont
  {Tanimura}},\ }\href@noop {} {\bibfield  {journal} {\bibinfo  {journal}
  {Physical Review A}\ }\textbf {\bibinfo {volume} {41}},\ \bibinfo {pages}
  {15} (\bibinfo {year} {1990})}\BibitemShut {NoStop}%
\bibitem [{\citenamefont {Tanimura}(2020)}]{Tanimura2020}%
  \BibitemOpen
  \bibfield  {author} {\bibinfo {author} {\bibfnamefont {Y.}~\bibnamefont
  {Tanimura}},\ }\href@noop {} {\bibfield  {journal} {\bibinfo  {journal}
  {Journal of Chemical Physics}\ }\textbf {\bibinfo {volume} {153}} (\bibinfo
  {year} {2020})}\BibitemShut {NoStop}%
\bibitem [{\citenamefont {Lambert}\ \emph {et~al.}(2023)\citenamefont
  {Lambert}, \citenamefont {Raheja}, \citenamefont {Cross}, \citenamefont
  {Menczel}, \citenamefont {Ahmed}, \citenamefont {Pitchford}, \citenamefont
  {Burgarth},\ and\ \citenamefont {Nori}}]{Lambert2023}%
  \BibitemOpen
  \bibfield  {author} {\bibinfo {author} {\bibfnamefont {N.}~\bibnamefont
  {Lambert}}, \bibinfo {author} {\bibfnamefont {T.}~\bibnamefont {Raheja}},
  \bibinfo {author} {\bibfnamefont {S.}~\bibnamefont {Cross}}, \bibinfo
  {author} {\bibfnamefont {P.}~\bibnamefont {Menczel}}, \bibinfo {author}
  {\bibfnamefont {S.}~\bibnamefont {Ahmed}}, \bibinfo {author} {\bibfnamefont
  {A.}~\bibnamefont {Pitchford}}, \bibinfo {author} {\bibfnamefont
  {D.}~\bibnamefont {Burgarth}}, \ and\ \bibinfo {author} {\bibfnamefont
  {F.}~\bibnamefont {Nori}},\ }\href@noop {} {\bibfield  {journal} {\bibinfo
  {journal} {Physical Review Research}\ }\textbf {\bibinfo {volume} {5}},\
  \bibinfo {pages} {013181} (\bibinfo {year} {2023})}\BibitemShut {NoStop}%
\bibitem [{\citenamefont {Langreth}\ and\ \citenamefont
  {Nordlander}(1991)}]{Langreth1991}%
  \BibitemOpen
  \bibfield  {author} {\bibinfo {author} {\bibfnamefont {D.~C.}\ \bibnamefont
  {Langreth}}\ and\ \bibinfo {author} {\bibfnamefont {P.}~\bibnamefont
  {Nordlander}},\ }\href@noop {} {\bibfield  {journal} {\bibinfo  {journal}
  {PHYSICAL REVIEW B}\ }\textbf {\bibinfo {volume} {43}},\ \bibinfo {pages}
  {2541} (\bibinfo {year} {1991})}\BibitemShut {NoStop}%
\bibitem [{\citenamefont {Wingreen}\ \emph {et~al.}(1989)\citenamefont
  {Wingreen}, \citenamefont {Jacobsen},\ and\ \citenamefont
  {Wilkins}}]{Wingreen1989}%
  \BibitemOpen
  \bibfield  {author} {\bibinfo {author} {\bibfnamefont {N.~S.}\ \bibnamefont
  {Wingreen}}, \bibinfo {author} {\bibfnamefont {K.~W.}\ \bibnamefont
  {Jacobsen}}, \ and\ \bibinfo {author} {\bibfnamefont {J.~W.}\ \bibnamefont
  {Wilkins}},\ }\href@noop {} {\bibfield  {journal} {\bibinfo  {journal}
  {Physical Review B}\ }\textbf {\bibinfo {volume} {40}},\ \bibinfo {pages}
  {11834} (\bibinfo {year} {1989})}\BibitemShut {NoStop}%
\bibitem [{\citenamefont {Dou}\ and\ \citenamefont {Subotnik}(2018)}]{Dou2018}%
  \BibitemOpen
  \bibfield  {author} {\bibinfo {author} {\bibfnamefont {W.}~\bibnamefont
  {Dou}}\ and\ \bibinfo {author} {\bibfnamefont {J.~E.}\ \bibnamefont
  {Subotnik}},\ }\href@noop {} {\bibfield  {journal} {\bibinfo  {journal}
  {Journal of Chemical Physics}\ }\textbf {\bibinfo {volume} {148}} (\bibinfo
  {year} {2018})}\BibitemShut {NoStop}%
\bibitem [{\citenamefont {Brako}\ and\ \citenamefont
  {Newns}(1981{\natexlab{b}})}]{Brako1981JPC}%
  \BibitemOpen
  \bibfield  {author} {\bibinfo {author} {\bibfnamefont {R.}~\bibnamefont
  {Brako}}\ and\ \bibinfo {author} {\bibfnamefont {D.~M.}\ \bibnamefont
  {Newns}},\ }\href {http://iopscience.iop.org/0022-3719/14/21/023} {\bibfield
  {journal} {\bibinfo  {journal} {J. Phys. C: Solid State Phys}\ }\textbf
  {\bibinfo {volume} {14}},\ \bibinfo {pages} {3065} (\bibinfo {year}
  {1981}{\natexlab{b}})}\BibitemShut {NoStop}%
\bibitem [{\citenamefont {Sch\"{o}nhammer}\ and\ \citenamefont
  {Gunnarsson}(1980)}]{Gunnarsson1980}%
  \BibitemOpen
  \bibfield  {author} {\bibinfo {author} {\bibfnamefont {K.}~\bibnamefont
  {Sch\"{o}nhammer}}\ and\ \bibinfo {author} {\bibfnamefont {O.}~\bibnamefont
  {Gunnarsson}},\ }\href@noop {} {\bibfield  {journal} {\bibinfo  {journal}
  {Physical Review B}\ }\textbf {\bibinfo {volume} {22}},\ \bibinfo {pages}
  {1629} (\bibinfo {year} {1980})}\BibitemShut {NoStop}%
\bibitem [{\citenamefont {Sch\"{o}nhammer}(1981)}]{Schornhammer1981}%
  \BibitemOpen
  \bibfield  {author} {\bibinfo {author} {\bibfnamefont {K.}~\bibnamefont
  {Sch\"{o}nhammer}},\ }\href@noop {} {\bibfield  {journal} {\bibinfo
  {journal} {Z. Phys. B-Condensed Matter}\ }\textbf {\bibinfo {volume} {45}},\
  \bibinfo {pages} {23} (\bibinfo {year} {1981})}\BibitemShut {NoStop}%
\bibitem [{\citenamefont {Kuznetsov}(1989)}]{Kuznetsov1989}%
  \BibitemOpen
  \bibfield  {author} {\bibinfo {author} {\bibfnamefont {A.~M.}\ \bibnamefont
  {Kuznetsov}},\ }\href@noop {} {\emph {\bibinfo {title} {Physical Chemistry:
  An Andvanced Treatise}}},\ edited by\ \bibinfo {editor} {\bibfnamefont
  {B.~E.}\ \bibnamefont {Conway}}, \bibinfo {editor} {\bibfnamefont {R.~J.}\
  \bibnamefont {White}}, \ and\ \bibinfo {editor} {\bibfnamefont {J.~O.}\
  \bibnamefont {Bockris}},\ Vol.~\bibinfo {volume} {20}\ (\bibinfo  {publisher}
  {Plenum},\ \bibinfo {year} {1989})\BibitemShut {NoStop}%
\bibitem [{\citenamefont {Schmickler}(1996)}]{Schmickler1996}%
  \BibitemOpen
  \bibfield  {author} {\bibinfo {author} {\bibfnamefont {W.}~\bibnamefont
  {Schmickler}},\ }\href@noop {} {\emph {\bibinfo {title} {Interfacial
  Electrochemistry}}}\ (\bibinfo  {publisher} {Oxford University},\ \bibinfo
  {year} {1996})\BibitemShut {NoStop}%
\bibitem [{\citenamefont {Santos}\ \emph {et~al.}(2006)\citenamefont {Santos},
  \citenamefont {Koper},\ and\ \citenamefont {Schmickler}}]{Santos2006}%
  \BibitemOpen
  \bibfield  {author} {\bibinfo {author} {\bibfnamefont {E.}~\bibnamefont
  {Santos}}, \bibinfo {author} {\bibfnamefont {M.~T.}\ \bibnamefont {Koper}}, \
  and\ \bibinfo {author} {\bibfnamefont {W.}~\bibnamefont {Schmickler}},\
  }\href@noop {} {\bibfield  {journal} {\bibinfo  {journal} {Chemical Physics
  Letters}\ }\textbf {\bibinfo {volume} {419}},\ \bibinfo {pages} {421}
  (\bibinfo {year} {2006})}\BibitemShut {NoStop}%
\bibitem [{\citenamefont {Santos}\ \emph {et~al.}(2008)\citenamefont {Santos},
  \citenamefont {Koper},\ and\ \citenamefont {Schmickler}}]{Santos2008}%
  \BibitemOpen
  \bibfield  {author} {\bibinfo {author} {\bibfnamefont {E.}~\bibnamefont
  {Santos}}, \bibinfo {author} {\bibfnamefont {M.~T.}\ \bibnamefont {Koper}}, \
  and\ \bibinfo {author} {\bibfnamefont {W.}~\bibnamefont {Schmickler}},\
  }\href@noop {} {\bibfield  {journal} {\bibinfo  {journal} {Chemical Physics}\
  }\textbf {\bibinfo {volume} {344}},\ \bibinfo {pages} {195} (\bibinfo {year}
  {2008})}\BibitemShut {NoStop}%
\bibitem [{\citenamefont {Nakanishi}\ \emph {et~al.}(1988)\citenamefont
  {Nakanishi}, \citenamefont {Kasai},\ and\ \citenamefont
  {Okiji}}]{Nakanishi1988}%
  \BibitemOpen
  \bibfield  {author} {\bibinfo {author} {\bibfnamefont {H.}~\bibnamefont
  {Nakanishi}}, \bibinfo {author} {\bibfnamefont {H.}~\bibnamefont {Kasai}}, \
  and\ \bibinfo {author} {\bibfnamefont {A.}~\bibnamefont {Okiji}},\
  }\href@noop {} {\bibfield  {journal} {\bibinfo  {journal} {Surface Science}\
  }\textbf {\bibinfo {volume} {197}},\ \bibinfo {pages} {515} (\bibinfo {year}
  {1988})}\BibitemShut {NoStop}%
\bibitem [{\citenamefont {Hamelin}\ and\ \citenamefont
  {Weaver}(1987)}]{Hamelin1987}%
  \BibitemOpen
  \bibfield  {author} {\bibinfo {author} {\bibfnamefont {A.}~\bibnamefont
  {Hamelin}}\ and\ \bibinfo {author} {\bibfnamefont {M.~J.}\ \bibnamefont
  {Weaver}},\ }\href@noop {} {\bibfield  {journal} {\bibinfo  {journal} {J.
  Electroanal. Chem.}\ }\textbf {\bibinfo {volume} {223}},\ \bibinfo {pages}
  {171} (\bibinfo {year} {1987})}\BibitemShut {NoStop}%
\bibitem [{\citenamefont {Schmickler}(1993)}]{Schmickler1993}%
  \BibitemOpen
  \bibfield  {author} {\bibinfo {author} {\bibfnamefont {W.}~\bibnamefont
  {Schmickler}},\ }\href@noop {} {\bibfield  {journal} {\bibinfo  {journal}
  {Surface Science}\ }\textbf {\bibinfo {volume} {295}},\ \bibinfo {pages} {43}
  (\bibinfo {year} {1993})}\BibitemShut {NoStop}%
\end{thebibliography}%

\end{document}